# Magnetic Amplification at $Yb^{3+}$ "Designer Defects" in the van der Waals Ferromagnet, $CrI_3$

Kimo Pressler, Thom J. Snoeren, Kelly M. Walsh, Daniel R. Gamelin*

*Department of Chemistry, University of Washington, Seattle, WA 98195, United States*

Email: *gamelin@uw.edu*

**Abstract.** The two-dimensional (2D) van der Waals ferromagnet $CrI_3$ has been doped with the magnetic optical impurity $Yb^{3+}$ to yield materials that display sharp multi-line $Yb^{3+}$ photoluminescence (PL) controlled by the magnetism of $CrI_3$. Magneto-PL shows that $Yb^{3+}$ magnetization is pinned to the magnetization of $CrI_3$. An effective internal field of ~10 T at $Yb^{3+}$ is estimated, attributed to strong in-plane $Yb^{3+}$-$Cr^{3+}$ superexchange coupling. The anomalously low energy of $Yb^{3+}$ PL in $CrI_3$ reflects relatively high $Yb^{3+}$-$I^-$ covalency, contributing to $Yb^{3+}$-$Cr^{3+}$ superexchange coupling. The $Yb^{3+}$ PL energy and linewidth both reveal the effects of spontaneous zero-field $CrI_3$ magnetic ordering *within* 2D layers below $T_C$, despite the absence of net magnetization in multilayer samples. These results illustrate the use of optical impurities as "designer defects" to introduce unique functionality to 2D magnets.

Defects have the power to transform the physical properties of crystals, imparting new and potentially useful functionalities from conductivity to quantum photon emission.[1-6] In magnetic materials, defects can strongly affect spin-wave propagation, magnetic domain-wall propagation, skyrmion dynamics, and magnetic vortex pinning.[7-9] Recently, the layered van der Waals ferromagnet $CrI_3$ has emerged as a promising platform for exploring strongly correlated spin physics, magnetic proximity effects, and next-generation spin-based device architectures in the two-dimensional (2D) limit,[10-14] but the potential to expand $CrI_3$ functionality through introduction of defects remains untapped. Here, we report that doping $CrI_3$ with $Yb^{3+}$ as a "designer point defect" transforms its normally broad and featureless *d-d* photoluminescence (PL) into narrow-line sensitized *f-f* emission, without compromising its attractive magnetic properties. We further show that $Yb^{3+}$ in $CrI_3$ experiences a large internal effective field that makes it extremely sensitive to small external magnetic fields. Using this property, we demonstrate magnetically saturated circular polarization of $Yb^{3+}$ emission at anomalously small applied fields. Strikingly, the internal effective field also transmits magnetic information to $Yb^{3+}$ even in the absence of any applied field, making $Yb^{3+}$ a unique embedded luminescent probe of spontaneous zero-field magnetic ordering within the 2D monolayers of bulk $CrI_3$. These discoveries establish optical impurity doping as an effective strategy for expanding the functionality of 2D magnets, with potential ramifications for both basic science and future spin-photonic technologies.

$CrI_3$ has become a model system for exploring magnetic exchange in 2D van der Waals structures,[10-14] stimulated by recent discoveries of Ising-like hard ferromagnetism in exfoliated monolayer $CrI_3$ and layer- and stacking-dependent magnetism in multi-layer $CrI_3$.[15,16] Layering $CrI_3$ with non-magnetic 2D materials introduces magnetic functionality to the non-magnetic material *via* inter-layer exchange coupling, allowing magnetic manipulation of properties such as $WSe_2$ valley polarization and valley Zeeman splittings.[17] Extension from few to many (bulk) layers preserves the strong Ising-like intralayer ferromagnetic ordering, but facile motion of domain walls unblocks demagnetization.[18] Despite its rich magnetic properties, $CrI_3$ itself has not garnered much attention as an optical material. Bulk $CrI_3$ has been investigated for its very large Kerr and Faraday rotation strengths in relation to optical isolators and associated technologies.[19,20] PL of bulk $CrI_3$ has apparently not been reported, and few-layer $CrI_3$ shows[17] only the very broad *d-d* PL characteristic of weak-field pseudo-octahedral $Cr^{3+}$.[21] Circular polarization of this *d-d* PL was used to probe the magnetism of few-layer $CrI_3$,[17] but the emission's breadth limits its further utility for fundamental studies or in spin-photonics, stimulating efforts to narrow the band *via* cavity coupling.[22] Doping $CrI_3$ with optically active impurities has also not been reported, either in bulk or exfoliated samples.

To investigate *intra*layer "proximity" effects resulting from magnetic exchange coupling, we have prepared $CrI_3$ doped with luminescent and spin-bearing $Yb^{3+}$ ions. Large-diameter single-crystal flakes of $CrI_3$ were prepared by chemical vapor transport. $Yb^{3+}$ was introduced by adding Yb(0) to the precursor mix. The $Yb^{3+}$ concentration in the resulting $Yb^{3+}$:$CrI_3$ crystals is controllable, and samples with up to ~5% $Yb^{3+}$ (cation mole fraction, $[Yb^{3+}]/([Cr^{3+}]+[Yb^{3+}])$) are described here. Further experimental details are provided in the Supporting Information (SI). Figure 1a shows a photograph of representative $Yb^{3+}$:$CrI_3$ flakes in their growth tube. The flakes are between 5 and 10 mm across, with typical thicknesses of 5-20 μm (see SI). Figure 1b plots XRD data collected on undoped and 4.9% $Yb^{3+}$-doped $CrI_3$ single-crystal flakes using a powder diffractometer. Only (00*l*) peaks are observed, corresponding to the interlayer lattice spacing and reflecting the flake's alignment. Figure 1c highlights the shift to smaller angle of the 001 peak upon doping. From fitting the XRD peak positions of undoped and 4.9% $Yb^{3+}$-doped $CrI_3$ samples, the interlayer lattice parameter was found to increase 0.24% from 6.996 ± 0.002 to 7.013 ± 0.002 Å, attributed to the larger ionic radius of $Yb^{3+}$ than $Cr^{3+}$ (87 *vs* 62 pm, respectively) (see SI). These data suggest that the local strain of doping is relieved by distorting the lattice along its softest dimension, as expected. Substitutional incorporation of $Yb^{3+}$ at the $Cr^{3+}$ site is verified by single-crystal XRD measurements (see SI), which also show the increased interlayer spacing. The single-crystal data show no detectable electron density between layers, ruling out $Yb^{3+}$ intercalation.

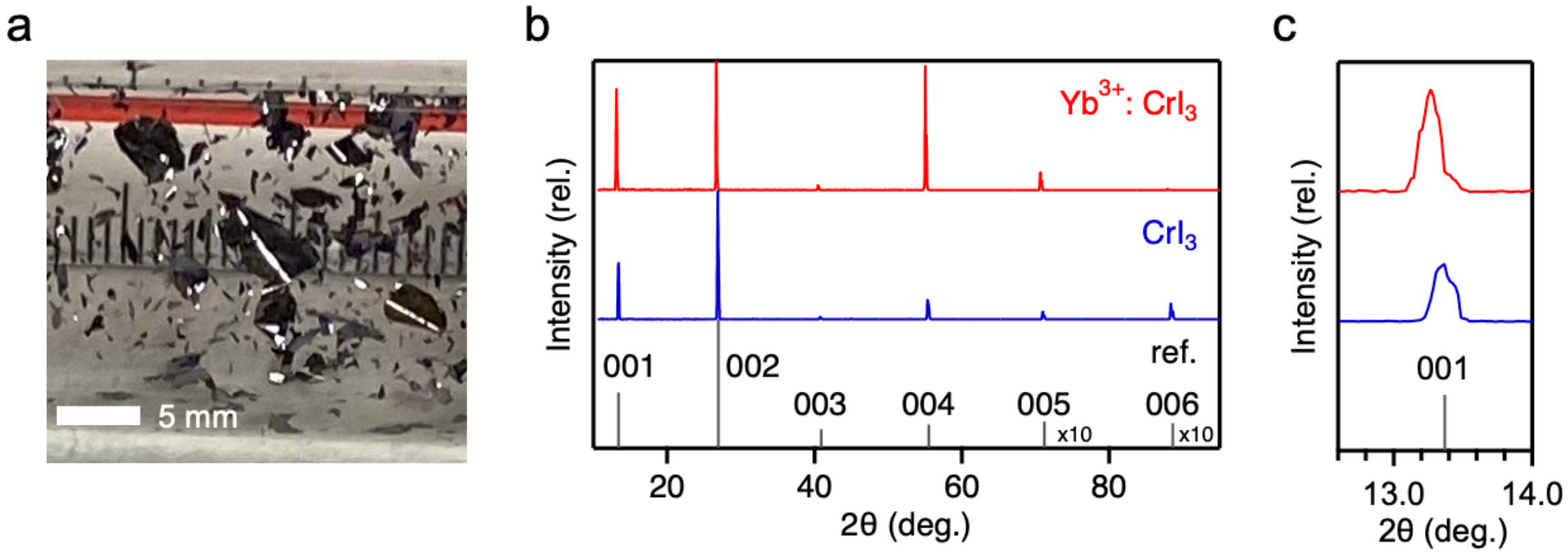

**Figure 1. (a)** Photograph of 4.9% $Yb^{3+}$:$CrI_3$ crystals prepared by chemical vapor transport. The scale bar shows 5 mm. All experiments were performed on individual single-crystal flakes from such a reaction tube. **(b)** XRD data collected on undoped and $Yb^{3+}$-doped $CrI_3$ single crystals using a powder diffractometer. Only (00*l*) peaks are observed, indicating an oriented sample. Reference peaks for *c*-oriented $CrI_3$ diffraction are included (black, ICSD Coll. Code 251654). **(c)** Magnified view of the 001 reflection for the same samples, displaying an increase in the interlayer lattice spacing upon $Yb^{3+}$ doping. The $x$ axis in (c) was determined as described in the SI.

Figure 2a plots the PL spectra of $CrI_3$ and $Yb^{3+}$:$CrI_3$ single flakes measured at several temperatures between 4 and 200 K. The $CrI_3$ spectrum broadens and decreases in intensity with increasing temperature, eventually reaching only 7.5% of its 4 K intensity at 200 K (see SI). Although the broadening to higher energies is expected from thermal hot bands, the broadening to lower energies is abnormal and suggests an additional feature. Upon introduction of $Yb^{3+}$, the broad featureless *d-d* emission of $Cr^{3+}$ disappears and is replaced by a series of sharp *f-f* transitions of $Yb^{3+}$ around 1.15 eV. Assignment of the PL fine structure is discussed later. In some samples, $Yb^{3+}$ doping also reveals another broad emission band centered at ~0.95 eV, which is responsible for the red tail of the $CrI_3$ PL here and in some literature spectra. This feature has been traced to $Ni^{2+}$ impurities (<0.4%) found in some Cr(0) precursors, and it can be mostly eliminated by using 5N Cr(0) precursors (Fig. 2a, bottom). The $Yb^{3+}$ PL is not influenced by this $Ni^{2+}$ impurity (see SI).

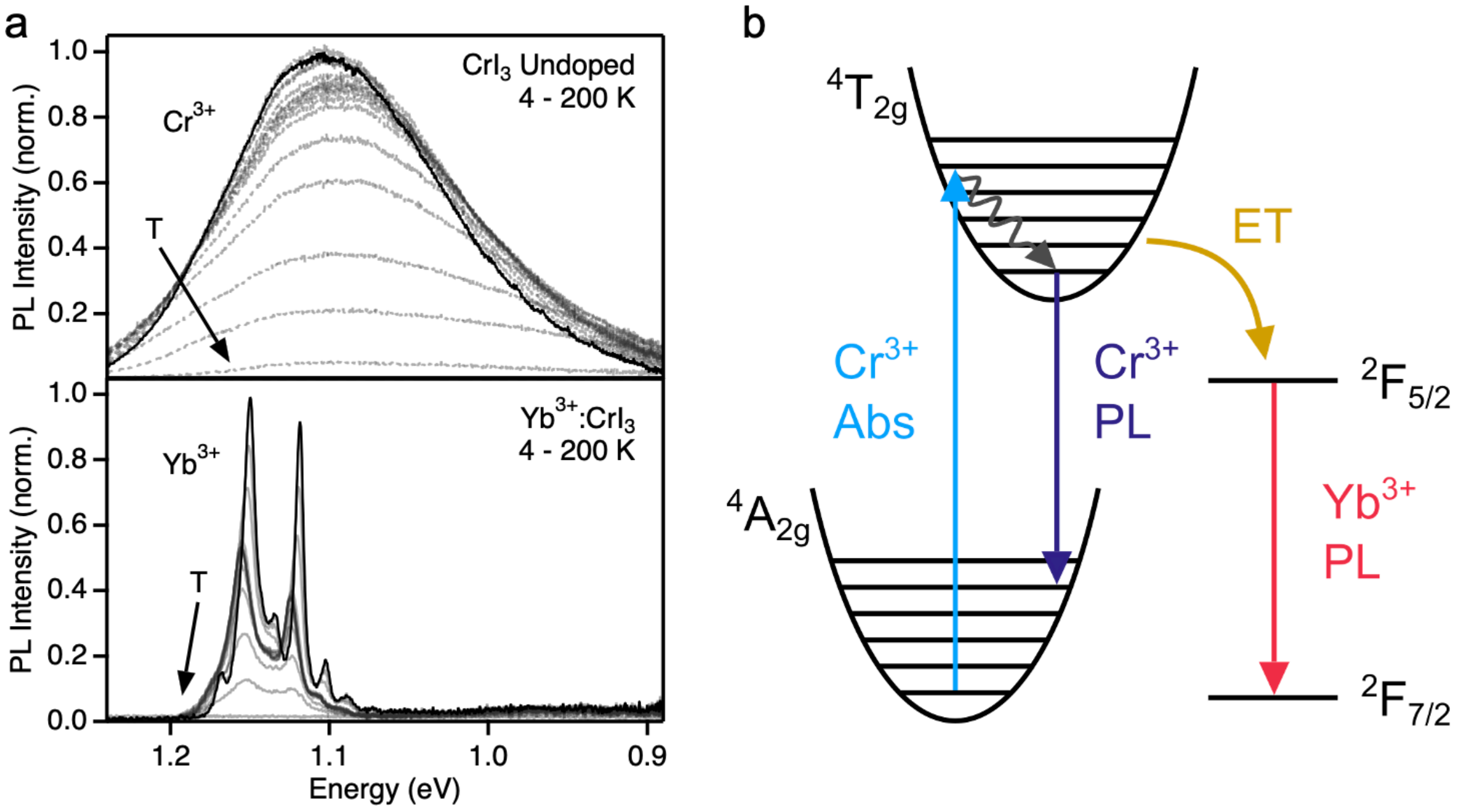

**Figure 2. (a)** Variable-temperature PL spectra of $CrI_3$ (top) and 4.9% $Yb^{3+}$:$CrI_3$ (bottom), measured from 4 to 200 K under 1.88 eV CW excitation at 4 mW/cm$^2$. **(b)** Single-configurational-coordinate diagram ($A_{1g}$ coordinate) describing vibronic broadening of the absorption and luminescence bands associated with transitions between the $^4A_{2g}$ and $^4T_{2g}$ ligand-field states of pseudo-octahedral $Cr^{3+}$. In $Yb^{3+}$-doped $CrI_3$, energy transfer from the $Cr^{3+}$ $^4T_{2g}$ excited state to $Yb^{3+}$ yields sensitized $^2F_{5/2} \rightarrow {}^2F_{7/2}$ *f-f* luminescence.

Figure 2b illustrates the photophysics of $Yb^{3+}$:$CrI_3$ schematically. The lowest-energy excited state of $CrI_3$ is the $Cr^{3+}$ $^4T_{2g}$ ligand-field state, involving excitation of a $t_{2g}$ electron into a σ-antibonding $e_g$ orbital (in idealized $O_h$ symmetry). The resulting change in equilibrium geometry is described by the single-configurational-coordinate (SCC) diagram of Fig. 2b, which illustrates the totally symmetric distortion coordinate. This $^4T_{2g}$ excited state also distorts along a symmetry-breaking Jahn-Teller coordinate (not illustrated).[21] These distortions lead to extensive vibronic progressions in the absorption and PL spectra associated with this transition, and cause a large PL Stokes shift. Doping $CrI_3$ with $Yb^{3+}$ introduces a set of $^2F_{5/2}$ states just below the $Cr^{3+}$ $^4T_{2g}$ excited state, favorably positioned for efficient $Cr^{3+} \rightarrow Yb^{3+}$ energy transfer. At 4.9% $Yb^{3+}$ doping, the $Cr^{3+}$ $^4T_{2g}$ PL is entirely quenched and strong $Yb^{3+}$ $^2F_{5/2}$ emission is observed in its place (Fig. 2a). Because both $Cr^{3+}$ and $Yb^{3+}$ states are localized at single ions, energy migration within the $CrI_3$ lattice is required for this complete quenching. In undoped $CrI_3$, energy migration among equivalent $Cr^{3+}$ sites may occur but is not readily apparent. In $Yb^{3+}$:$CrI_3$, this energy migration is interrupted when energy is captured by $Yb^{3+}$ dopants. In 4.9% $Yb^{3+}$:$CrI_3$, the average $Cr^{3+}$ ion has only ~14% probability of having a neighboring $Yb^{3+}$, and ~50% probability of having at least one $Yb^{3+}$ within its first two cation shells. Energy must therefore migrate over at least a few lattice sites within the $^4T_{2g}$ lifetime to fully quench the $Cr^{3+}$ emission as observed in Fig. 2a.

Figure 3a shows the anticipated electronic structure of $Yb^{3+}$ in $CrI_3$. In the free ion, spin-orbit coupling splits the $^2F$ term into $^2F_{5/2}$ (excited) and $^2F_{7/2}$ (ground) states by an amount $\Delta E = 7/2\zeta$, where $\zeta = 361.8$ meV is the free-ion spin-orbit coupling constant.[23] In crystals, each of these states is further split by the crystal field. Figure 3b shows circularly polarized PL spectra of 4.9% $Yb^{3+}$:$CrI_3$ measured in a 0.5 T field applied parallel to the crystal's $c$ axis (*vide infra*). Three zero-phonon electronic origins are observed and assigned to the $\Gamma_8 \rightarrow \Gamma_6$, $\Gamma_8$, and $\Gamma_7$ transitions anticipated from Fig. 3a using idealized $O_h$ notation. The actual cation site symmetry in $CrI_3$ is lower (Fig. 3a, right),[24] but the expected low-symmetry splitting of the $\Gamma_8$ origin is not clearly identifiable. The $\Gamma_6$ peak is broad with observable structure on its high-energy shoulder, thus making the precise energy of this origin unclear within ~20 $cm^{-1}$ (~2.5 meV). Analysis of these PL energies within the Angular Overlap Model (AOM)[25] reproduces the $^2F_{7/2}$ splittings well, predicting a $^2F_{5/2}$ splitting of ~34 meV and splittings of the two $\Gamma_8$ levels by <0.5 meV each (see SI). Additional satellite features are observed ~127 $cm^{-1}$ (15.7 meV) below the $\Gamma_8$ and $\Gamma_7$ electronic origins and assigned as phonon sidebands. Raman spectra show a totally symmetric lattice breathing mode of $CrI_3$ at this energy ($\nu$ = 127 $cm^{-1}$).[26]

A striking aspect of this $Yb^{3+}$:$CrI_3$ PL is its very low energy relative to other $Yb^{3+}$ PL. This energy is primarily determined by spin-orbit coupling (Fig. 3a). $Yb^{3+}$ spin-orbit coupling can be reduced from that in the free ion by covalent expansion of the *f*-electron wavefunctions (nephelauxetic effect),[27,28] but *f*-orbital covalency in trivalent lanthanides is typically very small and this effect is usually considered negligible at ambient pressure. A survey of $Yb^{3+}$-doped crystals shows that the energy gap between $Yb^{3+}$ $^2F_{5/2}$ and $^2F_{7/2}$ barycenters remains very near the free-ion value of $\Delta E$ ~ 1.266 eV across doped oxide, fluoride, chloride, bromide, sulfide, and phosphide lattices (see SI).[29-33] We note that we have been unable to find *any* reports of PL from other $Yb^{3+}$-doped iodide crystals, perhaps because $Yb^{3+}$ is easily reduced to $Yb^{2+}$ under common iodide crystal-growth conditions. $Yb^{3+}$:$CrI_3$ deviates from this typical behavior substantially: $\Delta E$ is only ~1.163 eV, or ~9% smaller than in the free ion, representing the smallest spin-orbit coupling yet reported for $Yb^{3+}$. Covalency in $Yb^{3+}$:$CrI_3$ is certainly enhanced by the large ionic radius and polarizability of the iodides, but this consideration alone likely cannot explain the anomaly. The atomic spin-orbit coupling of I is also much greater than those of other common ligands for $Yb^{3+}$, and should contribute to the spectroscopic spin-orbit splitting *via* covalency. Furthermore, the large ionic radius of $Yb^{3+}$ compared to $Cr^{3+}$ means that $Yb^{3+}$ experiences an internal pressure imposed by the surrounding lattice, which may also increase covalency. Importantly, $Yb^{3+}$-$I^-$ covalency is essential for strong $Yb^{3+}$-$Cr^{3+}$ superexchange coupling.

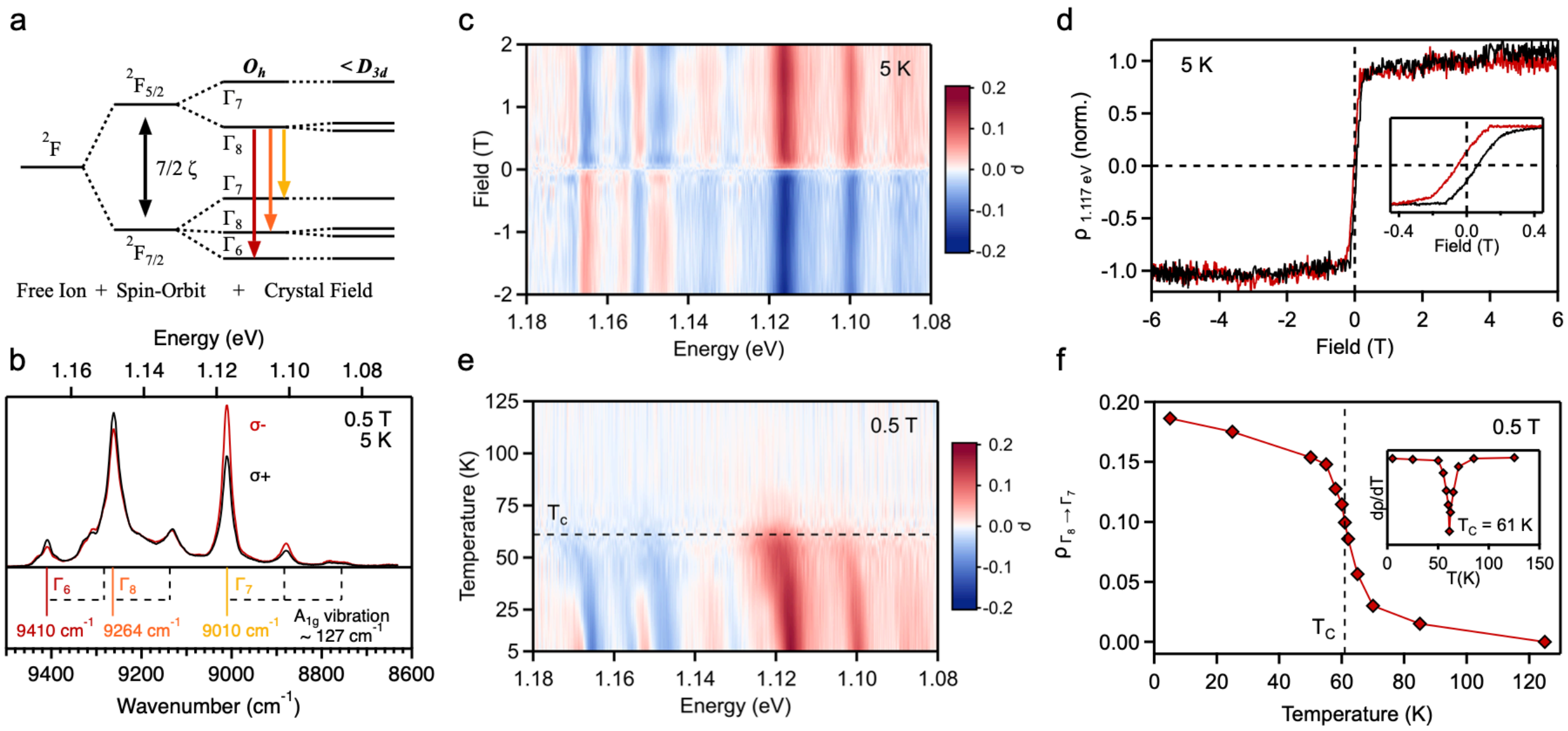

**Figure 3. (a)** Splitting of the $Yb^{3+}$ free-ion $^2F$ term due to spin-orbit ($\zeta$) and crystal-field ($O_h$, $<D_{3d}$) interactions. The colored down arrows indicate the three crystal-field transitions anticipated in the low-temperature PL spectrum in the idealized $O_h$ site symmetry. The actual site symmetry is reduced to $<D_{3d}$, *e.g.*, to $C_2$, splitting each $\Gamma_8$ level into two Kramers doublets. **(b)** Magnetic circularly polarized luminescence (MCPL) spectra of 4.9% $Yb^{3+}$:$CrI_3$ measured at 5 K with an applied magnetic field of 0.5 T. The $\sigma^-$ (red) and $\sigma^+$ (black) spectra were collected using unpolarized 1.88 eV CW excitation at 40 mW/cm$^2$ and have different amplitudes. The three electronic origins in idealized $O_h$ symmetry are indicated below the spectra, assigned to the $\Gamma_8 \rightarrow \Gamma_6$, $\Gamma_8$, and $\Gamma_7$ transitions illustrated in panel (a). The dashed black lines indicate vibronic sidebands with a characteristic energy spacing of ~127 cm$^{-1}$ (15.7 meV), consistent with the $A_{1g}$ lattice mode of $CrI_3$. **(c)** False-color plot of the MCPL polarization ratio, $\rho = (\sigma^- - \sigma^+)/(\sigma^- + \sigma^+)$, for the full $Yb^{3+}$ PL spectrum, measured from -2 to +2 T at 5 K. **(d)** $\rho$ of the $\Gamma_8 \rightarrow \Gamma_7$ electronic origin (1.117 eV) plotted as a function of magnetic field from -6 to 6 T. The black (red) trace corresponds to the positive (negative) field sweep direction. Inset: Expanded plot of $\rho$ between -0.4 and +0.4 T, showing a coercive field of ~55 mT. For both field-sweep measurements, the sample was excited with linearly polarized 1.96 eV excitation, but with different powers (see Methods). **(e)** False-color plot of the polarization ratio *vs* temperature, measured at 0.5 T. The dashed black line indicates the Curie temperature of bulk $CrI_3$ ($T_C$ = 61 K). **(f)** Plot of the $\Gamma_8 \rightarrow \Gamma_7$ polarization ratio at the peak maximum measured at 0.5 T as a function of temperature. The red curve is a guide to the eye. Inset:

Derivative of $\rho$ as a function of temperature. The extracted Curie temperature is 61 K, indistinguishable from that of the undoped crystal.

From Fig. 3a, all features show circularly polarized PL, with the $\Gamma_8 \rightarrow \Gamma_7$ origin showing the greatest polarization ratio ($\rho = (\sigma^- - \sigma^+)/(\sigma^- + \sigma^+) = 19\%$). $\rho$ is independent of excitation power but its maximum value varies somewhat between samples (see SI). Figure 3c plots $\rho$ across the entire PL spectrum as a function of magnetic field. All $Yb^{3+}$ transitions are influenced by the applied field in the same way, consistent with all PL arising from the same excited state ($\Gamma_8$). Figure 3d plots $\rho$ for the $\Gamma_8 \rightarrow \Gamma_7$ peak as a function of applied field. $\rho$ increases rapidly at very low fields and saturates at only ~0.2 T. Increasing the field from 0.2 to 6.0 T does not change $\rho$ further, consistent with complete magnetization of $Yb^{3+}$ by 0.2 T. On an expanded scale, these data show a hysteresis with coercivity of ~55 mT, comparable to that found in magnetic measurements of bulk $CrI_3$.[18,34] We note that these $\rho$ values are generally small compared to those in cubic $Yb^{3+}$:InP (~70% at 10 T, 4.2 K),[33] possibly suggesting an in-plane or canted $Yb^{3+}$ anisotropy. Figure 3e summarizes the temperature dependence of $\rho$, measured at 0.5 T, and Fig. 3f highlights the temperature dependence for $\Gamma_8 \rightarrow \Gamma_7$ individually. All spectral features behave similarly, showing a pronounced drop in polarization at the Curie temperature of bulk $CrI_3$ (~61 K, see Fig. 3f, inset). These magneto-optical data agree well with magnetic susceptibility data (see SI), and both indicate that $Yb^{3+}$ doping causes no significant change in the magnetic characteristics of $CrI_3$ in these samples. This MCPL field and temperature dependence is highly unusual for $Yb^{3+}$, which generally shows simple paramagnetism of a pseudo-spin 1/2. For example, our AOM crystal-field analysis (see SI) predicts $g_{avg} \sim 2.7$ for the lowest $^2F_{7/2}$ Kramers doublet. Overall, the anomalous magnetism seen in the $Yb^{3+}$ MCPL reflects *magnetic* integration of $Yb^{3+}$ with ferromagnetic $CrI_3$.

Magnetic ordering was originally explained by Weiss in terms of a huge internal "molecular field"[35] exerted upon individual ions by their surrounding magnetic matrix, and this model provides a useful heuristic for estimating the effective field experienced by $Yb^{3+}$ within $CrI_3$. In this model, the effective field is given by the sum of external and molecular fields, as in eq 1.

$$H_{\mathrm{eff}} = H_{\mathrm{ext}} + H_{\mathrm{mol}} \tag{1}$$

In Fig. 3c,d, $CrI_3$ reaches magnetic saturation at very small $H_{\mathrm{ext}}$ (<0.2 T). At such low fields, $H_{\mathrm{ext}} \ll H_{\mathrm{mol}}$, and hence $H_{\mathrm{eff}} \sim H_{\mathrm{mol}}$. In the molecular-field model, $H_{\mathrm{mol}}$ in $CrI_3$ is given by eq 2,

$$H_{mol} = \frac{2zJ\langle S\rangle}{g\mu_B} \tag{2}$$

where, $J$ is the nearest-neighbor exchange coupling constant, $z = 3$ in $CrI_3$, $g$ is the Landé $g$ factor (2.00 for $Cr^{3+}$ in $CrI_3$), $\mu_B$ is the Bohr magneton, and $\langle S\rangle$ is the spin expectation value for $Cr^{3+}$ in $CrI_3$, whose absolute value equals 3/2 at saturation. $T_C$ in this model is determined by $J$ according to eq 3,

$$T_C = \frac{2zJS(S+1)}{3k_B} \tag{3}$$

where $S = 3/2$ for $Cr^{3+}$, and $k_B$ is the Boltzmann constant. From $T_C = 61$ K, eq 3 yields a value of $J = 0.70$ meV in $CrI_3$. Entering this $J$ value into eq 2 yields $H_{\mathrm{mol}}$ = ~54 T in $CrI_3$. $H_{\mathrm{mol}}$ is dominated by superexchange coupling, since dipolar contributions cannot account for the high $T_C$ of $CrI_3$.[36] For $Yb^{3+}$ in $CrI_3$, $J$ is reduced by the shielding of the 4*f* orbitals. $Cr^{3+}$(*d*)-$Yb^{3+}$(*f*) superexchange coupling has received relatively little experimental or theoretical attention,[37-39] but relevant experimental data are found in inelastic neutron scattering analyses of $Cs_3Yb_{1.8}Cr_{0.2}Br_9$, where $Yb^{3+}$-$Cr^{3+}$ exchange splittings are ~1/4 those for $Cr^{3+}$-$Cr^{3+}$.[37] This scaling factor is approximate because of the different lattice structure, but $Cs_3Yb_{1.8}Cr_{0.2}Br_9$ is the most similar halide-bridged $Yb^{3+}$-$Cr^{3+}$ system for which reliable exchange-coupling strengths could be found. This rough scaling reduces $H_{\mathrm{mol}}$ to ~14 T. Accounting for the larger $g$ value of $Yb^{3+}$ (~2.7, see SI), our best estimate is $H_{mol}$ ~ 10 T for $Yb^{3+}$ ions within $CrI_3$. Future spectroscopic measurements (*e.g.*, inelastic neutron scattering, Mössbauer, *etc.*) and calculations will be needed to refine this estimate, but the central conclusion drawn from both the experimental data and this analysis is clear: $Yb^{3+}$ magnetization in $Yb^{3+}$:$CrI_3$ is effectively pinned to the magnetic ordering of the $CrI_3$ lattice through strong $Yb^{3+}$-$Cr^{3+}$ superexchange coupling. The large $H_{mol}$ in $Yb^{3+}$:$CrI_3$ is attributable in large part to the $Yb^{3+}$-$I^-$ covalency discussed above. For comparison, exchange fields of 1.7 and ~1.1 T are reported for $Yb^{3+}$ in ferrimagnetic hexagonal $YbFeO_3$[40] and distorted orthorhombic $YbCrO_3$.[41] At these values, $Yb^{3+}$ magnetization is not pinned to the ordered $TM^{3+}$ spin sublattices.

A further remarkable aspect of $Yb^{3+}$:$CrI_3$ is that the effects of $H_{mol}$ are evident even at zero magnetic field ($H_{ext}$ = 0). Figure 4a plots zero-field $Yb^{3+}$ PL spectra as a function of temperature from 4 to 200 K. Viewing the data starting from high temperature, the peak positions appear nearly constant until roughly $T_C$. Below $T_C$, the peaks all shift to lower energy together. This redshift is also evident in Fig. 3e. Figure 4b highlights the temperature dependence of the $\Gamma_8 \rightarrow \Gamma_7$ transition energy. From 120 K to ~$T_C$, the transition energy increases gradually by only ~2 meV. Such temperature dependence has been variously modeled in terms of Raman scattering of non-resonant phonons or direct absorption/emission of phonons resonant with a crystal-field splitting.[42,43] For example, both models reproduce the $^2F_{7/2} \rightarrow {}^2F_{5/2}$ transition energies of $Yb^{3+}$:YAG well, whereas the resonant phonon model reproduces absorption linewidths marginally better.[43] As such, we apply the resonant phonon model here. The PL energies above $T_C$ are thus described by eq 4,[42,43]

$$E(T) = E_0 + \frac{\alpha_s}{e^{\Delta/k_B T} - 1} \qquad T > T_C \qquad (4)$$

where $E_0$ is the energy at 0 K, $\alpha_s$ describes the electron-phonon interaction strength, and Δ is the energy of the activating phonon mode, fixed at Δ = 127 $cm^{-1}$ (15.7 meV, Fig. 3b).

The solid curve in the high-temperature portion of Fig. 4b (>$T_C$) shows a fit to the high-temperature data using eq 4, floating $E_0$ and $\alpha_s$ and yielding best-fit values of 1.1242 eV and -6.3 meV, respectively. Eq 4 plateaus at $E_0$ in the limit of 0 K (dashed line < $T_C$ in Fig. 4b), but the experimental peak energy shows a discontinuity at $T_C$, dropping sharply and decreasing with decreasing temperature until reaching ~7 meV below $E_0$ in the low-temperature limit. With its link to $T_C$ and its characteristic curvature, this trend in $Yb^{3+}$ PL energy is associated with the spontaneous magnetization of individual $CrI_3$ monolayers, even though there is no net magnetization in these samples.

Spontaneous ferromagnetic ordering is classified as a second-order phase transition and, within the theory of universal scaling laws, is characterized by the order parameter $\beta$ shown in eq 5 describing the magnetization temperature dependence.[44]

$$M(T) = M_0\left(-\frac{T-T_c}{T_c}\right)^{\beta} \tag{5}$$

$M_0$ is the saturation moment per magnetic ion and equals 3.1 $\mu_B$ for $CrI_3$.[18] The precise value of $\beta$ depends on the underlying spin physics, but it is commonly around 1/3.[12] Previous examination of bulk $CrI_3$ found a critical exponent of $\beta$ = 0.284, between that expected from the 3D Ising model ($\beta$ = 0.325) and that of the tri-critical mean-field model ($\beta$ = 0.250).[34] Accordingly, the data in Fig. 4b below $T_C$ were simulated using eq 6 (sum of eq 4 and eq 5, with eq 4 parameters fixed by the high-temperature data). The scaling parameter ($\gamma$) in eq 6 relates magnetization to PL energy shift. The data are reproduced well using fixed values of $\beta$ = 1/3, $T_C$ = 60 K, and Δ = 127 $cm^{-1}$ (15.7 meV), with $\gamma$ as the only adjustable parameter. Relating eqs 5 and 6, these results indicate an $Yb^{3+}$ PL energy shift of -2.2 meV/$\mu_B$ during spontaneous $CrI_3$ intralayer magnetization. We stress that the zero-field PL data in Fig. 4 are not magnetic data, but highlight the strong influence of $CrI_3$ spontaneous magnetization on the $Yb^{3+}$ PL. Because $T_C$ in these samples is indistinguishable from that of bulk $CrI_3$ (Figs. 3f, S15), we tentatively attribute the small apparent broadening of the PL energy discontinuity around $T_C$ in Fig. 4b to additional PL hot bands that are not spectrally resolved.

$$E(T) = E_0 + \frac{\alpha_s}{e^{\Delta/k_B T} - 1} + \gamma\left(-\frac{T - T_c}{T_c}\right)^{\beta} \qquad T < T_C \qquad (6)$$

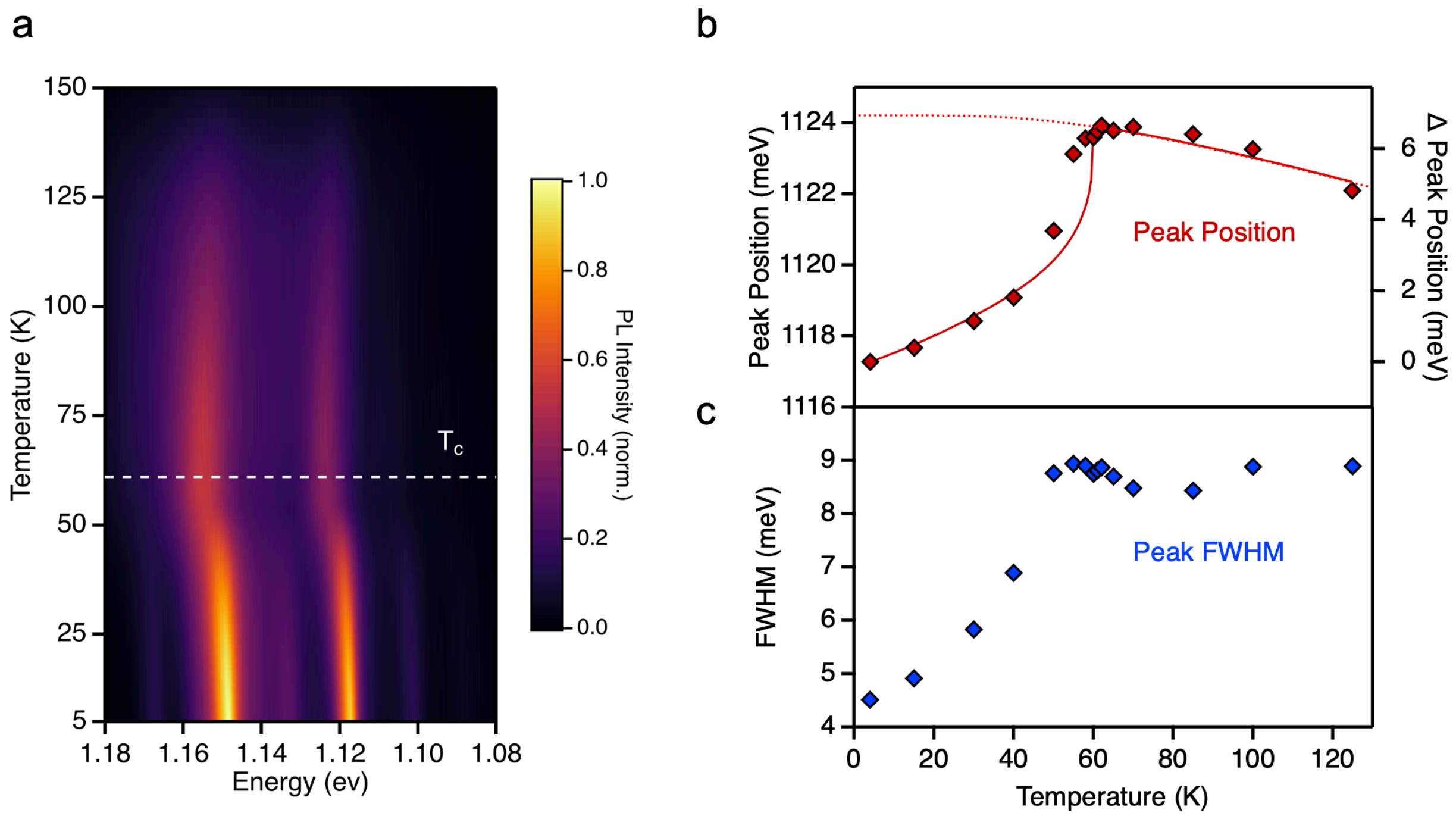

**Figure 4. (a)** False-color plot of the $Yb^{3+}$ PL intensities *vs* temperature measured for 4.9% $Yb^{3+}$:$CrI_3$ from 4 to 150 K at zero external magnetic field. The horizontal dashed line indicates $T_C$ = 61 K. **(b)** Peak position of the $\Gamma_8 \rightarrow \Gamma_7$ transition plotted *vs* temperature. The solid red curve shows the behavior predicted from the combination of resonant phonon interactions (eq 4) and spontaneous magnetization (below $T_C$, eq 6). The dashed red curve shows the behavior predicted from eq 4 alone below $T_C$. The solid curve was obtained using eqs 4 and 6 with fixed parameters of $\Delta$ = 127 $cm^{-1}$ (15.7 meV), $T_C$ = 60 K, and $\beta$ = 1/3, adjusting only the amplitude scaling. **(c)** Plot of the $\Gamma_8 \rightarrow \Gamma_7$ PL linewidth *vs* temperature, from the same VTPL measurements.

Figure 4c plots the temperature dependence of the $\Gamma_8 \rightarrow \Gamma_7$ linewidth (full-width-at-half-maximum, FWHM). These data show similar trends as observed in the peak energies of Fig. 4b. Below $T_C$, the FWHM decreases from ~9 meV to ~4.5 meV in the low-temperature limit, attributed to the reduction in spin disorder around $Yb^{3+}$. These data thus also reflect spontaneous magnetic ordering in monolayers of $CrI_3$. Although distinct low-energy shoulders are not resolved in these data, we hypothesize that the energy and linewidth changes below $T_C$ both ultimately stem from loss of hot-magnon sideband intensity as $CrI_3$ monolayers order magnetically.[45] It will be an interesting future direction to explore magnon coupling to *f-f* transitions in these and related doped 2D magnetic materials.

In summary, doping $Yb^{3+}$ into the 2D van der Waals ferromagnet $CrI_3$ transforms this material's PL from broad-band to sharp multi-line, while retaining its key magnetic functionality. The *f-f* PL of $Yb^{3+}$:$CrI_3$ is anomalously low in energy, reflecting relatively covalent $Yb^{3+}$-$I^-$ bonding. $Yb^{3+}$ magnetization is pinned to $CrI_3$ by strong superexchange interactions, which contribute an effective internal field of ~10 T that is greater than the field required for magnetic saturation of paramagnetic $Yb^{3+}$ and much greater than the field required for full $CrI_3$ magnetization at low temperature (~0.2 T). Flipping the magnetization of $CrI_3$ with a small external field thus also flips the $Yb^{3+}$ magnetization and inverts its PL circular polarization. Magnetic pinning is maintained up to the $T_C$ of $CrI_3$, but is rapidly lost above $T_C$. We further showed that the $Yb^{3+}$ PL energy and linewidth both sense this internal field even at zero applied field, mapping spontaneous *intra*layer magnetic ordering below $T_C$ despite the absence of net magnetization. Because each $Yb^{3+}$ ion is a local lattice defect within an individual $CrI_3$ monolayer, we expect these induced functionalities to persist down to the monolayer, prompting future studies on exfoliated $Yb^{3+}$:$CrI_3$ and associated stacked van der Waals heterostructures and layered devices. These results demonstrate the power of designer defects to add functionality to 2D magnetic materials, enrich their fundamental physics, and create new materials of potential utility for future spin-photonics applications.

**Acknowledgments.** Support of this project by the US NSF (DMR-1807394) is gratefully acknowledged. Initial stages of this work were performed as part of Programmable Quantum Materials, an Energy Frontier Research Center funded by the U.S. Department of Energy (DOE), Office of Science, Basic Energy Sciences (BES), under award DESC0019443. Additional support was received from the UW Clean Energy Institute (graduate fellowships to T.J.S. and K.M.W.). Part of this work was conducted at the Molecular Analysis Facility, a National Nanotechnology Coordinated Infrastructure (NNCI) site at the University of Washington that is supported in part by the National Science Foundation (NNCI-1542101 and NNCI-2025489), the University of Washington, the Molecular Engineering & Sciences Institute, the Clean Energy Institute, and the National Institutes of Health. The authors thank Dr. Werner Kaminsky and Paige M. Gannon for single-crystal XRD measurements, Dr. Xi Wang for assistance with optical microscope measurements, Prof. Jiun-Haw Chu and Dr. Zhaoyu Liu for VSM measurements, and Prof. Robert Glaum, Maximilian Jähnig, and Julia Spitz for provision of and assistance with the BonnMag code.

## Author Information

### Corresponding Author

**Daniel R. Gamelin** - *Department of Chemistry, University of Washington, Seattle, Washington 98195-1700, United States*; orcid.org/0000-0003-2888-9916; Email: gamelin@chem.washington.edu

### Authors

**Kimo Pressler** - *Department of Chemistry, University of Washington, Seattle, Washington 98195-1700, United States;* orcid.org/0000-0003-2788-1592

**Thom J. Snoeren** - *Department of Chemistry, University of Washington, Seattle, Washington 98195-1700, United States;* orcid.org/0000-0001-8055-3710

**Kelly M. Walsh** - *Department of Chemistry, University of Washington, Seattle, Washington 98195-1700, United States;* orcid.org/0000-0001-5349-8816

## Supporting Information

The Supporting Information is available free of charge at https://pubs.acs.org/doi/XXXX

Additional experimental details, including about sample preparation and characterization. Additional variable-temperature PL data, PL polarization *vs* magnetic field data, excitation-power-dependence data, results from $Yb^{3+}$ crystal-field calculations, and comparison of $Yb^{3+}$ crystal-field barycenter energies in various lattices (PDF).

**References**

(1) Cox, P. A., *Electronic Structure and Chemistry of Solids*. Oxford University Press: Oxford, 1987.

(2) Henderson, B.; Imbusch, G. F., *Optical Spectroscopy of Inorganic Solids*. Oxford University Press: Oxford, 1989.

(3) Kittel, C., *Introduction to Solid State Physics*. 8 ed.; Wiley: New York, 2004.

(4) Bassett, L. C.; Alkauskas, A.; Exarhos, A. L.; Fu, K.-M. C., Quantum defects by design. *Nanophotonics* **2019,** *8*, 1867-1888.

(5) Tran, T. T.; Elbadawi, C.; Totonjian, D.; Lobo, C. J.; Grosso, G.; Moon, H.; Englund, D. R.; Ford, M. J.; Aharonovich, I.; Toth, M., Robust Multicolor Single Photon Emission from Point Defects in Hexagonal Boron Nitride. *ACS Nano* **2016,** *10*, 7331-7338.

(6) Hong, J.; Jin, C.; Yuan, J.; Zhang, Z., Atomic Defects in Two-Dimensional Materials: From Single-Atom Spectroscopy to Functionalities in Opto-/Electronics, Nanomagnetism, and Catalysis. *Adv. Mater.* **2017,** *29*, 1606434.

(7) Cowley, R. A.; Buyers, W. J. L., The Properties of Defects in Magnetic Insulators. *Rev. Mod. Phys.* **1972,** *44*, 406-450.

(8) Wiesendanger, R., Nanoscale Magnetic Skyrmions in Metallic Films and Multilayers: A New Twist for Spintronics. *Nat. Rev. Mater.* **2016,** *1*, 16044.

(9) Lima Fernandes, I.; Bouaziz, J.; Blügel, S.; Lounis, S., Universality of Defect-Skyrmion Interaction Profiles. *Nat. Commun.* **2018,** *9*, 4395.

(10) Song, T.; Cai, X.; Tu, M. W.-Y.; Zhang, X.; Huang, B.; Wilson, N. P.; Seyler, K. L.; Zhu, L.; Taniguchi, T.; Watanabe, K.; McGuire, M. A.; Cobden, D. H.; Xiao, D.; Yao, W.; Xu, X., Giant Tunneling Magnetoresistance in Spin-Filter van der Waals Heterostructures. *Science* **2018,** *360*, 1214-1218.

(11) Huang, B.; McGuire, M. A.; May, A. F.; Xiao, D.; Jarillo-Herrero, P.; Xu, X., Emergent Phenomena and Proximity Effects in Two-Dimensional Magnets and Heterostructures. *Nat. Mater.* **2020,** *19*, 1276-1289.

(12) Gibertini, M.; Koperski, M.; Morpurgo, A. F.; Novoselov, K. S., Magnetic 2D materials and heterostructures. *Nat. Nanotech.* **2019,** *14*, 408-419.

(13) Gong, C.; Zhang, X., Two-dimensional magnetic crystals and emergent heterostructure devices. *Science* **2019,** *363*, eaav4450.

(14) Wang, Q. H.; Bedoya-Pinto, A.; Blei, M.; Dismukes, A. H.; Hamo, A.; Jenkins, S.; Koperski, M.; Liu, Y.; Sun, Q.-C.; Telford, E. J.; Kim, H. H.; Augustin, M.; Vool, U.; Yin, J.-X.; Li, L. H.; Falin, A.; Dean, C. R.; Casanova, F.; Evans, R. F. L.; Chshiev, M.; Mishchenko, A.; Petrovic, C.; He, R.; Zhao, L.; Tsen, A. W.; Gerardot, B. D.; Brotons-Gisbert, M.; Guguchia, Z.; Roy, X.; Tongay, S.; Wang, Z.; Hasan, M. Z.; Wrachtrup, J.; Yacoby, A.; Fert, A.; Parkin, S.; Novoselov, K. S.; Dai, P.; Balicas, L.; Santos, E. J. G., The Magnetic Genome of Two-Dimensional van der Waals Materials. *ACS Nano* **2022,** *16*, 6960-7079.

(15) Huang, B.; Clark, G.; Navarro-Moratalla, E.; Klein, D. R.; Cheng, R.; Seyler, K. L.; Zhong, D.; Schmidgall, E.; McGuire, M. A.; Cobden, D. H.; Yao, W.; Xiao, D.; Jarillo-Herrero, P.; Xu, X., Layer-Dependent Ferromagnetism in a van der Waals Crystal Down to the Monolayer Limit. *Nature* **2017,** *546*, 270-273.
(16) Sivadas, N.; Okamoto, S.; Xu, X.; Fennie, C. J.; Xiao, D., Stacking-Dependent Magnetism in Bilayer $CrI_3$. *Nano Lett.* **2018,** *18*, 7658-7664.
(17) Seyler, K. L.; Zhong, D.; Huang, B.; Linpeng, X.; Wilson, N. P.; Taniguchi, T.; Watanabe, K.; Yao, W.; Xiao, D.; McGuire, M. A.; Fu, K.-M. C.; Xu, X., Valley Manipulation by Optically Tuning the Magnetic Proximity Effect in $WSe_2/CrI_3$ Heterostructures. *Nano Lett.* **2018,** *18*, 3823-3828.
(18) McGuire, M. A.; Dixit, H.; Cooper, V. R.; Sales, B. C., Coupling of Crystal Structure and Magnetism in the Layered, Ferromagnetic Insulator $CrI_3$. *Chem. Mater.* **2015,** *27*, 612-620.
(19) Dillon, J. F.; Kamimura, H.; Remeika, J. P., Magneto-Optical Properties of Ferromagnetic Chromium Trihalides. *J. Phys. Chem. Solids* **1966,** *27*, 1531-1549.
(20) Suits, J., Faraday and Kerr Effects in Magnetic Compounds. *IEEE Trans. Mag.* **1972,** *8*, 95-105.
(21) Güdel, H. U.; Snellgrove, T. R., Jahn-Teller Effect in the $^4T_{2g}$ State of Chromium(III) in Dicesium Sodium Indium(III) Hexachloride. *Inorg. Chem.* **1978,** *17*, 1617-1620.
(22) Peng, B.; Chen, Z.; Li, Y.; Liu, Z.; Liang, D.; Deng, L., Multiwavelength Magnetic Coding of Helical Luminescence in Ferromagnetic 2D Layered $CrI_3$. *iScience* **2022,** *25*, 103623.
(23) Wyart, J.-F.; Tchang-Brillet, W.-Ü. L.; Spector, N.; Palmeri, P.; Quinet, P.; Biémont, E., Extended Analysis of the Spectrum of Triply-ionized Ytterbium (Yb IV) and Transition Probabilities. *Phys. Scripta* **2001,** *63*, 113-121.
(24) Georgescu, A. B.; Millis, A. J.; Rondinelli, J. M., Trigonal Symmetry Breaking and Its Electronic Effects in the Two-Dimensional Dihalides $MX_2$ and Trihalides $MX_3$. *Phys. Rev. B* **2022,** *105*, 245153.
(25) Bronova, A.; Bredow, T.; Glaum, R.; Riley, M. J.; Urland, W., BonnMag: Computer Program for Ligand-Field Analysis of $f^n$ Systems Within the Angular Overlap Model. *J. Comp. Chem.* **2018,** *39*, 176-186.
(26) Zhang, Y.; Wu, X.; Lyu, B.; Wu, M.; Zhao, S.; Chen, J.; Jia, M.; Zhang, C.; Wang, L.; Wang, X.; Chen, Y.; Mei, J.; Taniguchi, T.; Watanabe, K.; Yan, H.; Liu, Q.; Huang, L.; Zhao, Y.; Huang, M., Magnetic Order-Induced Polarization Anomaly of Raman Scattering in 2D Magnet $CrI_3$. *Nano Lett.* **2020,** *20*, 729-734.
(27) Al-Mobarak, R.; Warren, K. D., The Effect of Covalency on the Spin—Orbit Coupling Constant. *Chem. Phys. Lett.* **1973,** *21*, 513-516.
(28) Bungenstock, C.; Tröster, T.; Holzapfel, W. B., Effect of Pressure on Free-Ion and Crystal-Field Parameters of $Pr^{3+}$ in LOCl (L = La, Pr, Gd). *Phys. Rev. B* **2000,** *62*, 7945-7955.
(29) Schwartz, R. W., Electronic Structure of the Octahedral Hexachloroytterbate Ion. *Inorg. Chem.* **1977,** *16*, 1694-1698.
(30) Kanellakopulos, B.; Amberger, H. D.; Rosenbauer, G. G.; Fischer, R. D., Zur Elektronenstruktur hochsymmetrischer Verbindungen der Lanthanoiden und Actinoiden—V: Paramagnetische Suszeptibilität und elektronisches Raman-Spektrum von $Cs_2NaYb(III)Cl_6$. *J. Inorg. Nuc. Chem.* **1977,** *39*, 607-611.
(31) Tsujii, N.; Imanaka, Y.; Takamasu, T.; Kitazawa, H.; Kido, G., Photoluminescence of $Yb^{3+}$-Doped $CuInS_2$ Crystals in Magnetic Fields. *J. Appl. Phys.* **2001,** *89*, 2706-2710.

(32) Haumesser, P.-H.; Gaumé, R.; Viana, B.; Antic-Fidancev, E.; Vivien, D., Spectroscopic and Crystal-Field Analysis of New Yb-doped Laser Materials. *J. Phys.: Cond. Mat.* **2001,** *13*, 5427-5447.
(33) de Maat-Gersdorf, I. Spectroscopic Analysis of Erbium-Doped Silicon and Ytterbium Doped Indium Phosphide. University of Amsterdam, 2001.
(34) Liu, Y.; Petrovic, C., Three-Dimensional Magnetic Critical Behavior in $CrI_3$. *Phys. Rev. B* **2018,** *97*, 014420.
(35) Coey, J. M. D., *Magnetism and Magnetic Materials*. Cambridge University Press 2010.
(36) Lado, J. L.; Fernández-Rossier, J., On the Origin of Magnetic Anisotropy in Two Dimensional $CrI_3$. *2D Mater.* **2017,** *4*, 035002.
(37) Aebersold, M. A.; Güdel, H. U.; Hauser, A.; Furrer, A.; Blank, H.; Kahn, R., Exchange Interactions in Mixed $Yb^{3+}$-$Cr^{3+}$ and $Yb^{3+}$-$Ho^{3+}$ Dimers: An Inelastic-Neutron-Scattering Investigation of $Cs_3Yb_{1.8}Cr_{0.2}Br_9$ and $Cs_3Yb_{1.8}Ho_{0.2}Br_9$. *Phys. Rev. B* **1993,** *48*, 12723-12731.
(38) Mironov, V. S.; Chibotaru, L. F.; Ceulemans, A., Exchange Interaction in the $YbCrBr_9^{3-}$ Mixed Dimer: The Origin of a Strong $Yb^{3+}$-$Cr^{3+}$ Exchange Anisotropy. *Phys. Rev. B* **2003,** *67*, 014424.
(39) Atanasov, M.; Daul, C.; Güdel, H. U., Modelling of Anisotropic Exchange Coupling in Rare-Earth -Transition-Metal Pairs: Applications to $Yb^{3+}$-$Mn^{2+}$ and $Yb^{3+}$-$Cr^{3+}$ Halide Clusters and Implications to the Light Up-Conversion. In *Comp. Chem.: Rev. Current Trends*, World Scientific: 2005; Vol. 9, pp 153-194.
(40) Cao, S.; Sinha, K.; Zhang, X.; Zhang, X.; Wang, X.; Yin, Y.; N'Diaye, A. T.; Wang, J.; Keavney, D. J.; Paudel, T. R.; Liu, Y.; Cheng, X.; Tsymbal, E. Y.; Dowben, P. A.; Xu, X., Electronic Structure and Direct Observation of Ferrimagnetism in Multiferroic Hexagonal $YbFeO_3$. *Phys. Rev. B* **2017,** *95*, 224428.
(41) Dalal, B.; Sarkar, B.; Dev Ashok, V.; De, S. K., Evolution of Magnetic Properties and Exchange Interactions in Ru Doped $YbCrO_3$. *J. Phys.: Cond. Mat.* **2016,** *28*, 426001.
(42) Imbusch, G. F.; Yen, W. M.; Schawlow, A. L.; McCumber, D. E.; Sturge, M. D., Temperature Dependence of the Width and Position of the $^2E \rightarrow {}^4A_2$ Fluorescence Lines of $Cr^{3+}$ and $V^{2+}$ in MgO. *Phys. Rev.* **1964,** *133*, A1029-A1034.
(43) Böttger, T.; Thiel, C. W.; Cone, R. L.; Sun, Y.; Faraon, A., Optical Spectroscopy and Decoherence Studies of $Yb^{3+}$:YAG at 968 nm. *Phys. Rev. B* **2016,** *94*, 045134.
(44) Fisher, M. E., The Theory of Equilibrium Critical Phenomena. *Rep. Prog. Phys.* **1967,** *30*, 615-730.
(45) Bermudez, V. M.; McClure, D. S., Spectroscopic studies of the two-dimensional magnetic insulators chromium trichloride and chromium tribromide—II. *J. Phys. Chem. Solids* **1979,** *40*, 149-173.

**Table of Contents Graphic**

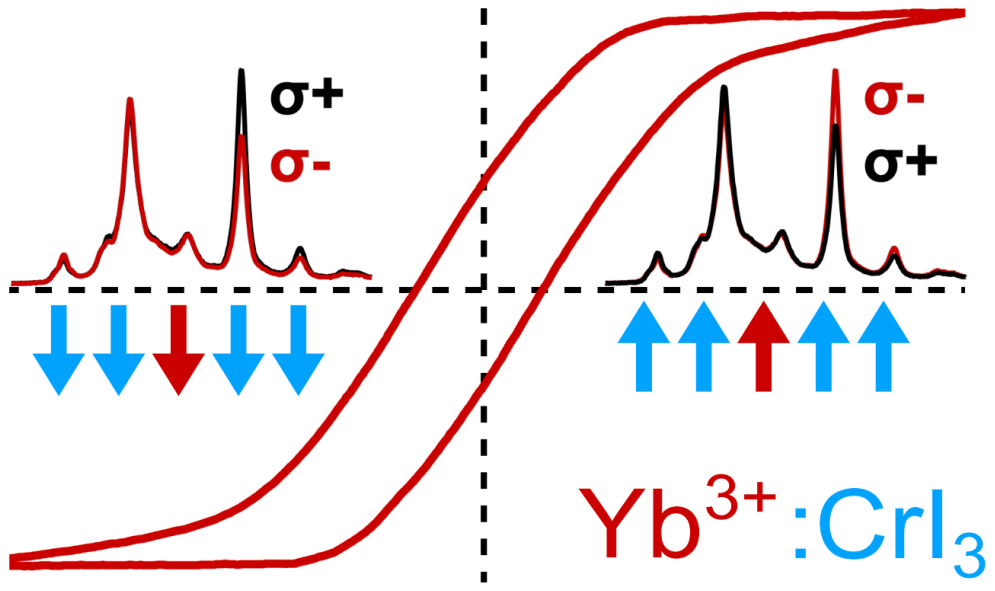

*Supporting Information for*

# Magnetic Amplification at $Yb^{3+}$ "Designer Defects" in the van der Waals Ferromagnet, $CrI_3$

Kimo Pressler, Thom J. Snoeren, Kelly M. Walsh, Daniel R. Gamelin*
*Department of Chemistry, University of Washington, Seattle, WA 98195, United States*
Email: *gamelin@uw.edu*

## Experimental Methods

**General Considerations.** All sample preparation and manipulation was performed in a glovebox under an atmosphere of purified dinitrogen.

**Chemicals.** Chromium metal powder (200 mesh, 99.94%, lot X15E028) was purchased from Alfa Aesar. According to the manufacturer's certificate of analysis, the majority of the impurity in this sample lot was Ni at 343 ppm. A chromium chip (99.995%, lot MKCH4484) was also purchased from Sigma Aldrich as a higher-purity Cr source. The Cr chip was ground to a powder using a mortar and pestle and used in an analogous manner as the powder precursor. $I_2$ (≥99.99%) was purchased from Sigma Aldrich. Ytterbium metal powder 40 mesh (99.9%) was purchased from BeanTown chemical. All chemicals were used as received without further purification.

**Synthesis of $CrI_3$ and $Yb^{3+}$-Doped $CrI_3$ Single Crystals.** Single crystals of the doped and undoped $CrI_3$ were grown by chemical vapor transport in a manner similar to that described in previous literature reports.[1] For undoped $CrI_3$, Cr(0) metal and $I_2$ were loaded as a 1:3 stochiometric ratio into a quartz tube and sealed under an evacuated atmosphere. For $Yb^{3+}$-doped $CrI_3$, additional Yb(0) metal was loaded along with the other starting materials. The quartz tubes were 15 cm long with inner and outer diameters of 14 and 16 mm, respectively. Sealed tubes were placed in an open-ended horizontal tube furnace with the starting materials in the hot zone set at 650 ˚C and the other end at a temperature of ca. 500 ˚C. Samples were heated for 5 days and then allowed to slowly cool to room temperature. Once cooled, the tubes were brought into a glove box and cracked open to yield shiny dark plate-like crystals that had formed at the cold end of the quartz tube. Elemental analysis of the $Yb^{3+}$-doped samples was performed by inductively coupled plasma mass spectrometry (ICP-MS) using a PerkinElmer NexION 2000B. Samples were prepared by digesting single crystals in concentrated nitric acid with sonication and then further diluted in ultrapure $H_2O$. $Yb^{3+}$ doping levels are reported as cation mole fraction, $[Yb^{3+}]/([Cr^{3+}]+[Yb^{3+}])$, in percentage, with an estimated uncertainty of ±0.1%. Crystal thickness was measured by mounting a representative flake to a glass slide using double-sided tape and imaging the flake with an optical microscope in a glovebox at various magnifications. The edge length was calculated in ImageJ[2] using known pixel resolutions.

**X-ray Diffraction (XRD) Characterization.** Samples were prepared for XRD on the powder diffractometer by placing single crystals onto silicon substrates and sealing under Kapton films to reduce exposure to air. Data were collected using a Bruker D8 Discover powder diffractometer with a high-efficiency IμS microfocus x-ray source for Cu Kα radiation (50 kV, 1 mA). For single-crystal XRD, a crystal measuring 0.10 x 0.05 x 0.01 $mm^3$ was mounted on a loop with oil. Data were collected at 263 K on a Bruker APEX II single-crystal X-ray diffractometer using Mo-radiation, equipped with a Miracol X-ray optical collimator. The data were integrated and scaled using SAINT, SADABS within the APEX2 software package by Bruker.[3] Solution by direct methods (SHELXT[4, 5] or SIR97[6, 7]) produced a complete heavy-atom phasing model consistent with the proposed structure. The structure was completed by difference Fourier synthesis with SHELXL.[8, 9] Scattering factors are from Waasmair and Kirfel.[10] All atoms were refined anisotropically by full-matrix least-squares.

Including intrinsic disorder, a least squares refinement optimization of the data yields the lattice structure that we report. From the 983 reflections collected covering the indices, $-8 \leq h \leq 8$, $-14 \leq k \leq 14$, $-8 \leq l \leq 8$, 518 reflections were found that were symmetry independent and an $R_1$ value of 0.0521 was obtained, indicating a good fit. $R_1$ is calculated as:

$$R_1 = \frac{\sum\left||F_{obs}| - |F_{calc}|\right|}{\sum|F_{obs}|}$$

There is no detectable electron density between layers, indicating that $Yb^{3+}$ does not intercalate between layers in $CrI_3$.

**Variable-Temperature Photoluminescence (VTPL).** Samples for VTPL measurements were prepared by placing a single crystal between two quartz disks and loading into a closed-cycle helium cryostat. PL spectra were collected by exciting the sample with a continuous-wave 660 nm (1.88 eV) diode at 4 mW/cm$^2$. Emission was collected and focused into a monochromator with a spectral bandwidth of 0.627 nm and detected by a Hamamatsu InGaAs/InP NIR photomultiplier tube, with signal recorded using a photon counter. Temperature was varied from 4 to 300 K, starting at low temperature. All spectra were corrected for instrument response.

**Magnetic Circularly Polarized Luminescence (MCPL).** Samples for MCPL measurements were prepared as single crystals placed between two quartz disks and loaded into a superconducting magneto-optical cryostat (Cryo-Industries SMC-1659 OVT) oriented in the Faraday configuration. For full-spectrum measurements at static fields, samples were excited with a 660 nm (1.88 eV) diode at approximately 40 mW/cm$^2$. For field-sweep measurements, samples were excited with a linearly polarized HeNe laser (632.8 nm/1.96 eV, 27 mW/cm$^2$ for -6 to +6 T scans, 55 mW/cm$^2$ for -0.4 to +0.4 T scans). No distinguishable difference was found in the either the PL spectra or variable-field data between the two excitation sources. For field-sweep measurements, the monochromator was centered at 1.117 eV with a 6 nm spectral bandwidth, and the signal was continuously monitored as the field was swept at a rate of 0.10 T/min and 0.45 T/min for the 0.4 T and 6 T scans, respectively. PL was collected along the magnetic-field axis and passed through a liquid-crystal variable retardation plate set at $\lambda/4$, followed by a linear polarizer to separate the left- and right-circularly polarized components. The PL was then focused into a fiber-optic cable and fed into a monochromator with a spectral bandwidth of 0.627 nm and detected by a Hamamatsu InGaAs/InP NIR photomultiplier, with signals recorded using a photon counter. Polarization ratios are defined as $\rho = (\sigma^- - \sigma^+)/(\sigma^- + \sigma^+) = (I_L - I_R)/(I_L + I_R) = \Delta I/I$, following the sign conventions outlined in Piepho and Schatz.[11]

**Magnetic Measurements.** Magnetic data on individual single-crystal flakes (Fig. 1) were collected using a Quantum Design PPMS DynaCool vibrating sample magnetometer (VSM). A flake was affixed to the end of a quartz paddle with varnish (VGE 7031). The paddle was then snapped into the VSM brass sample holder with another quartz paddle placed symmetrically above the sample. The weak background signal from the sample holder was removed in the data analysis. The sample was probed with the external field aligned perpendicular to the face of the crystal, and magnetization data were collected as a function of applied field and temperature. The masses of individual flakes are below 0.1 mg and could not be accurately measured, so the magnetic data are reported in units of emu.

**Ligand-field calculations within the Angular Overlap Model (AOM).** $Yb^{3+}$ ligand(crystal)-field energies and *g* factors were calculated using the BonnMag package.[12] Crystallographic data[13] on $CrI_3$ were used to create an $[YbI_6]^{3-}$ unit with reduced symmetry (point group $C_2$). Crystallographic parameters were not adjusted for size differences between $Cr^{3+}$ and $Yb^{3+}$. The electronic structure of $Yb^{3+}$ was calculated using the spin-orbit coupling parameter $\zeta$ as well as AOM parameters $e_\sigma$ and $e_\pi$ to describe σ and π interactions with the $I^-$ ligands, respectively. The value for $e_\pi$ was taken to be isotropic. The Slater-Condon-Shortley (SCS) parameters $F_2$, $F_4$, and $F_6$ were taken to be 0, as is typically the case for $Yb^{3+}$ ($4f^{13}$ configuration). The Stevens orbital reduction factor *k* was taken to be equal to 1.0. Increasing (decreasing) $\zeta$ while keeping all other parameters constant results in an increase (decrease) in all transition energies while retaining peak splitting energies. Adjusting $e_\sigma$ or $e_\pi$ alters the relative energies of the peaks but maintains the barycenters.

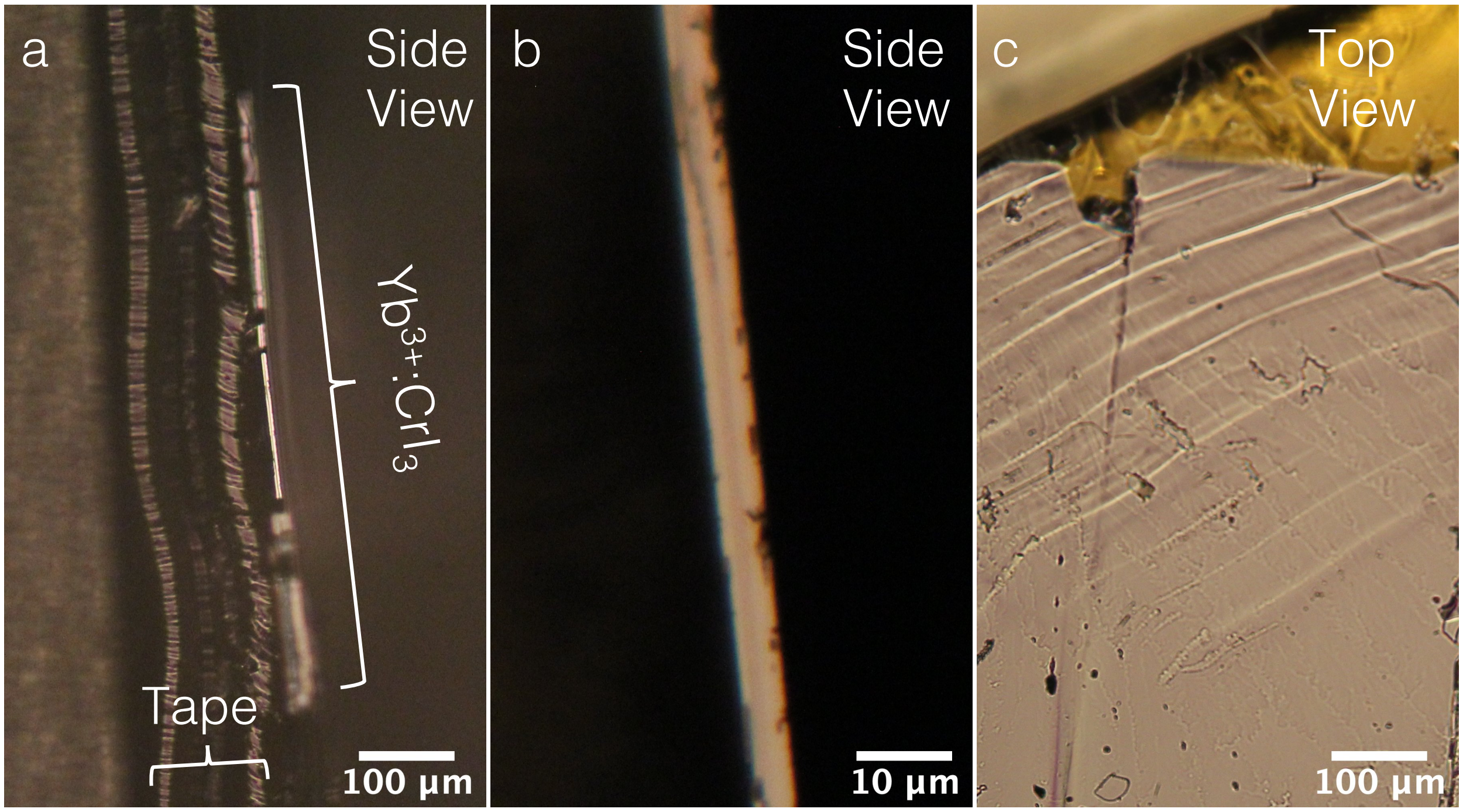

**Figure S1.** Images of an individual $Yb^{3+}$:$CrI_3$ single-crystal flake under an optical microscope at various magnification levels, viewing the flake's **(a,b)** edge, and **(c)** face. The flake thickness is estimated to be 5.1 ± 0.3 µm.

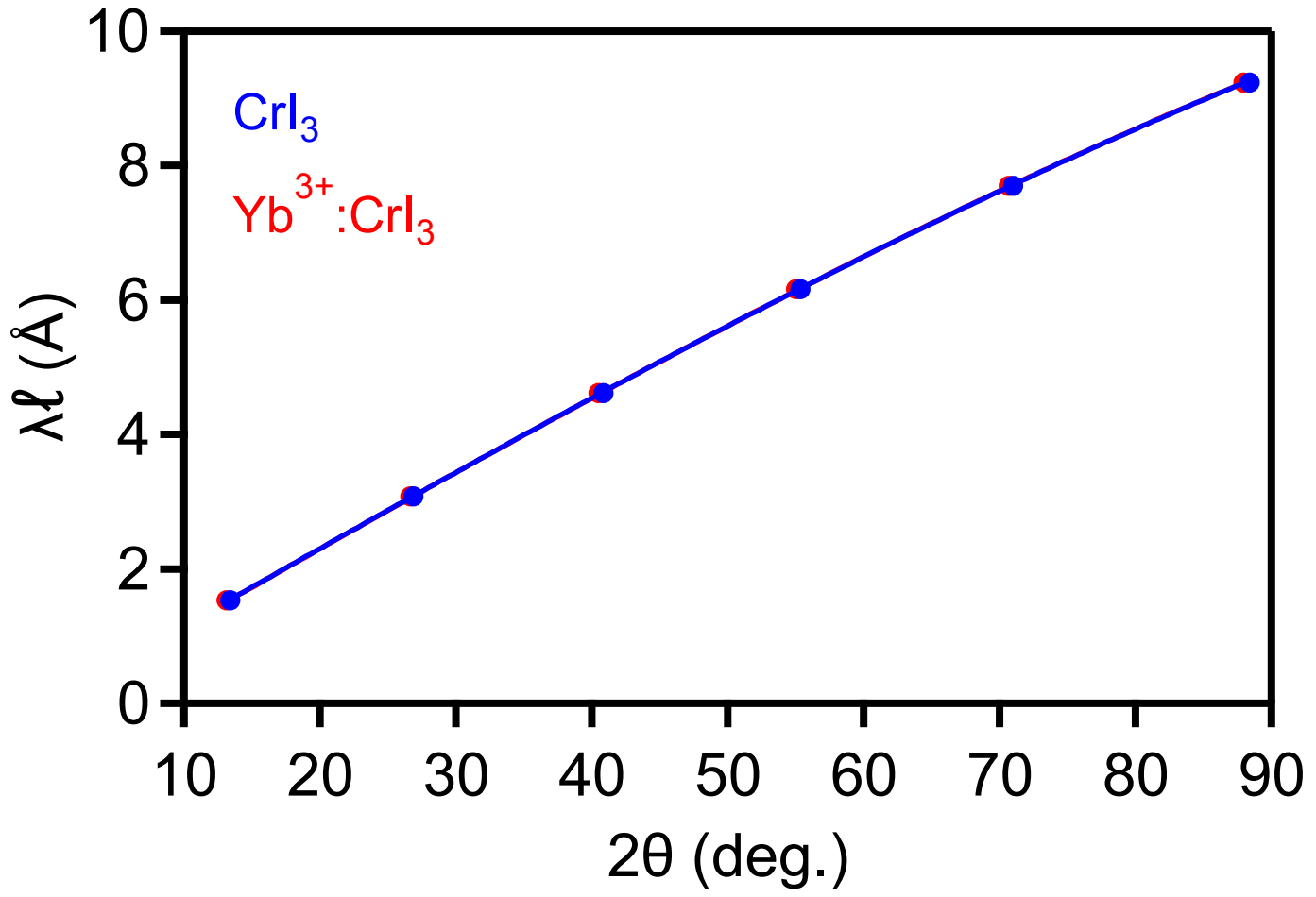

**Figure S2.** Analysis of XRD reflections collected using a powder diffractometer for 4.9% $Yb^{3+}$-doped and undoped $CrI_3$ single-crystal flakes (same data as shown in Fig. 1bc). Using the method described by Jesche,[14] the lattice parameter *c* for oriented single crystals with a monoclinic space group can be extracted from XRD data from a powder diffractometer using the following equation:

$$2c \cdot \sin\beta \cdot \sin\left(\theta - S\frac{\cos\theta}{2}\right) = \lambda\ell$$

Here, β is the obtuse angle in the monoclinic unit cell (108.507° for $CrI_3$), λ is the x-ray wavelength (Cu, 1.5406 Å), ℓ is the Miller index of each reflection in the XRD spectrum and $S\frac{\cos\theta}{2}$ is a correction factor related to the displacement of the x-ray focal plane relative to the sample surface. Plotting 2θ values of the peak maxima *vs* λℓ, the data can be fit using the equation above. For fitting, β and θ were taken in radians. By this method, the *c* lattice parameters were found to be 6.996 ± 0.002 and 7.013 ± 0.002 Å for the undoped and doped samples, respectively. From the lattice parameter *c*, the position of the (00ℓ) powder diffractometer XRD peaks for a monoclinic single crystal can be calculated using the following equation:

$$2\theta = 2\sin^{-1}\left(\frac{\lambda}{2\sin\beta}\frac{\ell}{c}\right)$$

The zero-shift in 2θ was determined by adding an offset to the experimental 2θ values and adjusting the offset to minimize the difference between experimental and calculated peak positions across all peaks in the XRD spectrum. This offset accounts for the measurement discrepancy due to the thickness of the single crystals displacing the x-ray focal plane. For $CrI_3$, a zero-shift of -0.015° was found, contrasted to a zero-shift of +0.164° for $Yb^{3+}$-doped $CrI_3$. The displacement-corrected XRD spectra are shown in Fig. 1c in the main text.

**Table S1. Single-crystal X-ray diffraction data for 2.5% $Yb^{3+}$:$CrI_3$ measured at 263 K, compared to literature data for $CrI_3$.**

| | $Yb^{3+}$:$CrI_3$ | $CrI_3$ (250 K, ref. [13]) |
|---|---|---|
| Space group | *C2/m* | *C2/m* |
| | | |
| *a* | 6.86 Å | 6.87 Å |
| *b* | 11.89 Å | 11.89 Å |
| *c* | 6.99 Å | 6.98 Å |
| *α* | 90.0° | 90.0° |
| *β* | 108.7° | 108.5° |
| *γ* | 90.0° | 90.0° |
| | | |
| $[(Yb/Cr) – Cr]_{avg}$ | 3.96 Å | 3.96 Å |
| $[(Yb/Cr) – I]_{avg}$ | 2.72 Å | 2.72 Å |
| $[(Yb/Cr) – I – (Yb/Cr)]_{avg}$ | 93.3° | 93.6° |
| $[I – (Yb/Cr) – I]_{avg}$ | 86.8° | 86.9° |

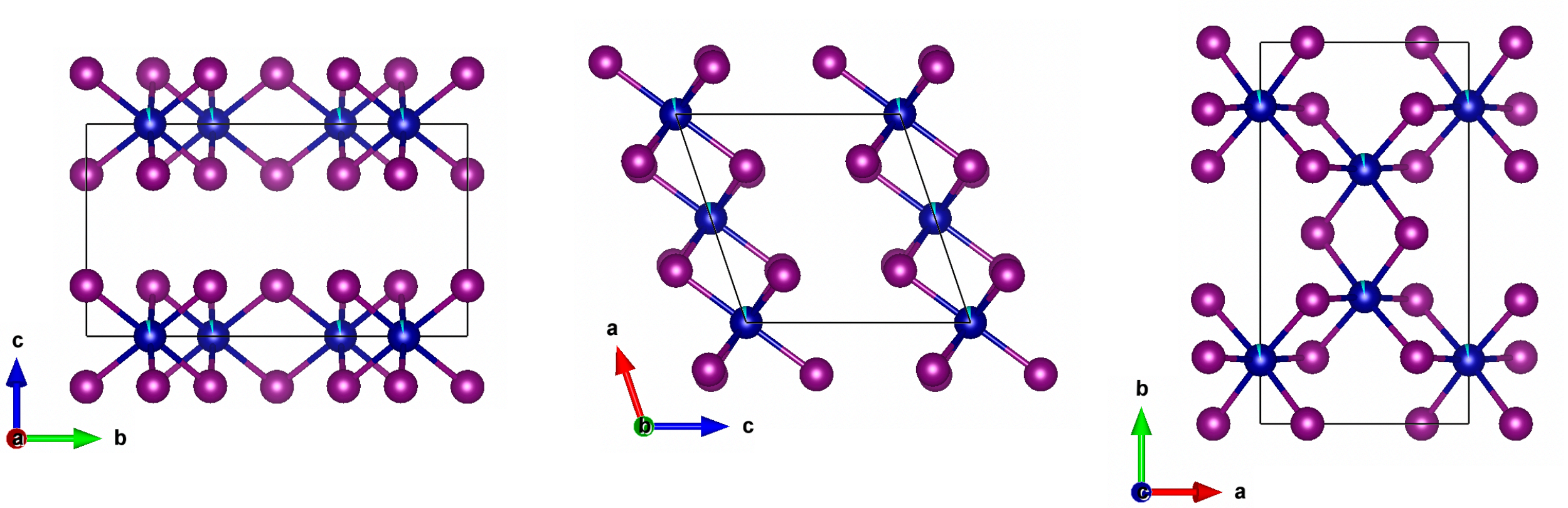

**Figure S3.** Visualization of the experimental room-temperature single-crystal XRD structure as viewed along the *a*, *b*, and *c* principal axes (left to right). $Yb^{3+}$ (cyan) is found to substitute for $Cr^{3+}$ (blue) in the edge-sharing octahedra formed by $I^-$ (purple) anions. No excess electron density is observed between layers. Intralayer disorder is observed. The structure refines to the expected high-temperature *C2/m* monoclinic symmetry. Some intralayer disorder was observed (not shown).

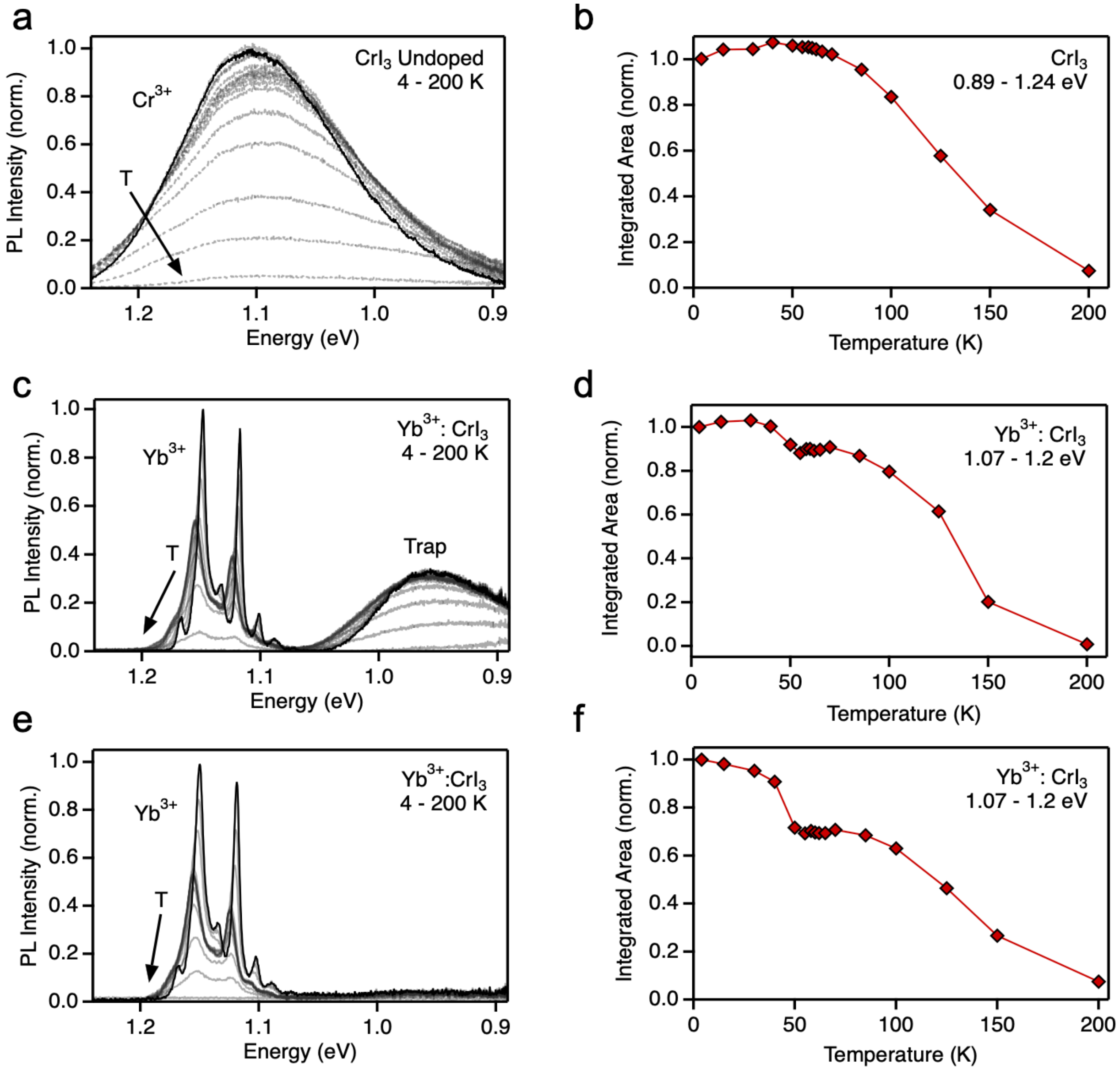

**Figure S4. (a)** Variable-temperature PL spectra of $CrI_3$ measured from 4 to 200 K under 1.88 eV CW excitation (from Fig. 2 of the main text). **(b)** Scatter plot depicting total integrated area of the $CrI_3$ PL from panel (a). The 200 K intensity is 7.5% that of the 4 K value. **(c)** Variable-temperature PL spectra of 4.9% $Yb^{3+}$:$CrI_3$ measured from 4 to 200 K under 1.88 eV CW excitation (from Fig. 2 of the main text). **(d)** Scatter plot depicting total integrated area of the $Yb^{3+}$ PL from panel (c). The 200 K intensity is 0.8% that of the 4 K value. **(e)** Variable-temperature PL spectra of 5.0% $Yb^{3+}$:$CrI_3$ measured from 4 to 200 K under 1.88 eV CW excitation (from Fig. 2 of the main text). **(f)** Scatter plot depicting total integrated area of the $Yb^{3+}$ PL from panel (e). The 200 K intensity is 7.5% that of the 4 K value. Note that a second, broad "trap" PL band is observed at ~0.98 eV in samples made from Cr metal powder precursor (99.94%, panel (c)) but not in samples made from Cr chip precursor (99.995%, panel (e)). Ni is the primary impurity in the powder precursor (see Methods), and Ni is detected in this $CrI_3$ sample at 0.4% cation mole fraction. $Ni^{2+}$ $^3A_{2g} \rightarrow {}^3T_{2g}$ absorption in $NiI_2$ and $Ni^{2+}$:$CdI_2$ is centered around 0.93 eV,[15] and the broad "trap" PL band in panel (c) is thus tentatively attributed to $Ni^{2+}$ impurities in $CrI_3$.

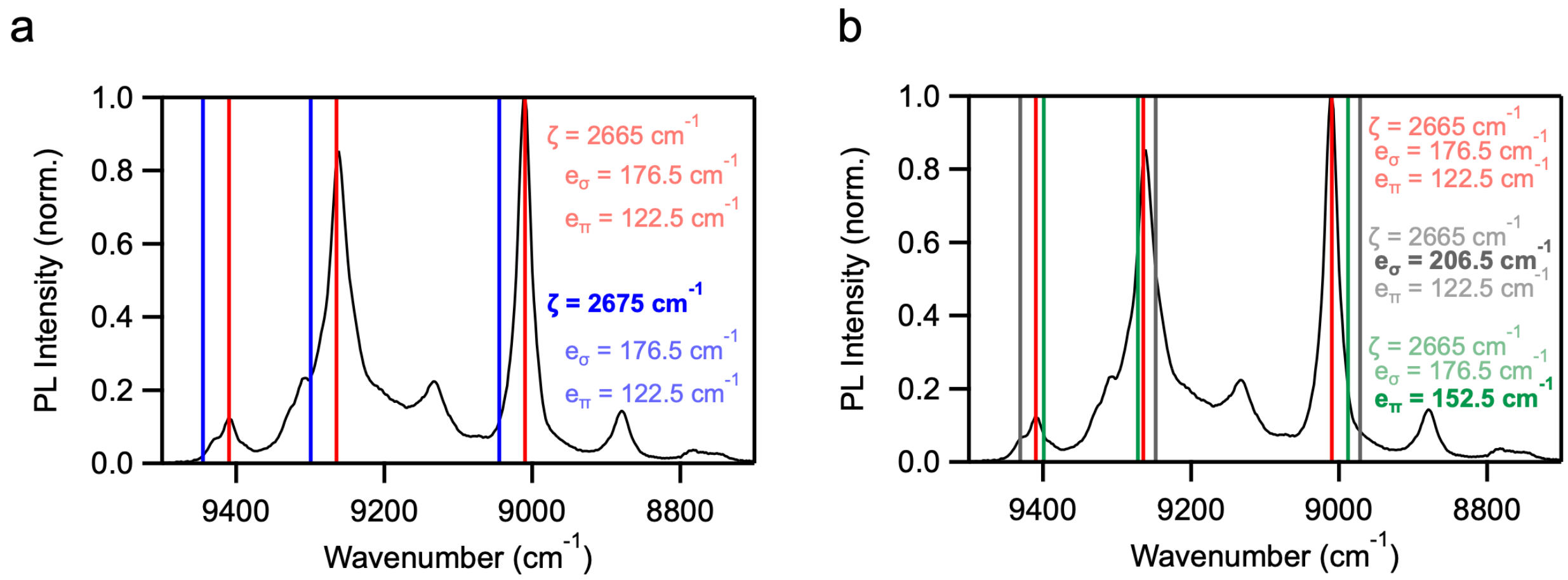

**Figure S5.** Comparison of the 5 K experimental data and calculated (AOM) *f-f* PL transition energies for 4.9% $Yb^{3+}$:$CrI_3$. A best fit to the experimental PL data resulted in the following values: $\zeta = 2665$ cm$^{-1}$ (330.4 meV), $e_\sigma = 176.5$ cm$^{-1}$ (21.9 meV), $e_\pi = 122.5$ cm$^{-1}$ (15.2 meV). The calculated transition energies using these parameters are shown as the vertical red lines in both panels. **(a)** Comparison of calculated transition energies obtained by changing from $\zeta = 2665$ cm$^{-1}$ (red) to $\zeta = 2675$ cm$^{-1}$ (blue), with all other parameters constant to the best-fit (red). **(b)** Comparison of calculated transition energies obtained by individually changing the values of $e_\sigma$ and $e_\pi$. The gray traces show the effect of changing from $e_\sigma = 176.5$ cm$^{-1}$ (red) to $e_\sigma = 206.5$ cm$^{-1}$ with all other parameters constant to the best fit (red). The green traces show the effect of changing from $e_\pi = 122.5$ cm$^{-1}$ (red) to $e_\pi = 152.5$ cm$^{-1}$ with all other parameters constant to the best fit (red). From the best-fit parameters, $g$ is anisotropic ($g_1 = 2.672$, $g_2 = 2.686$, $g_3 = 2.642$) and an average ground-state $g$ value of ~2.7 is predicted.

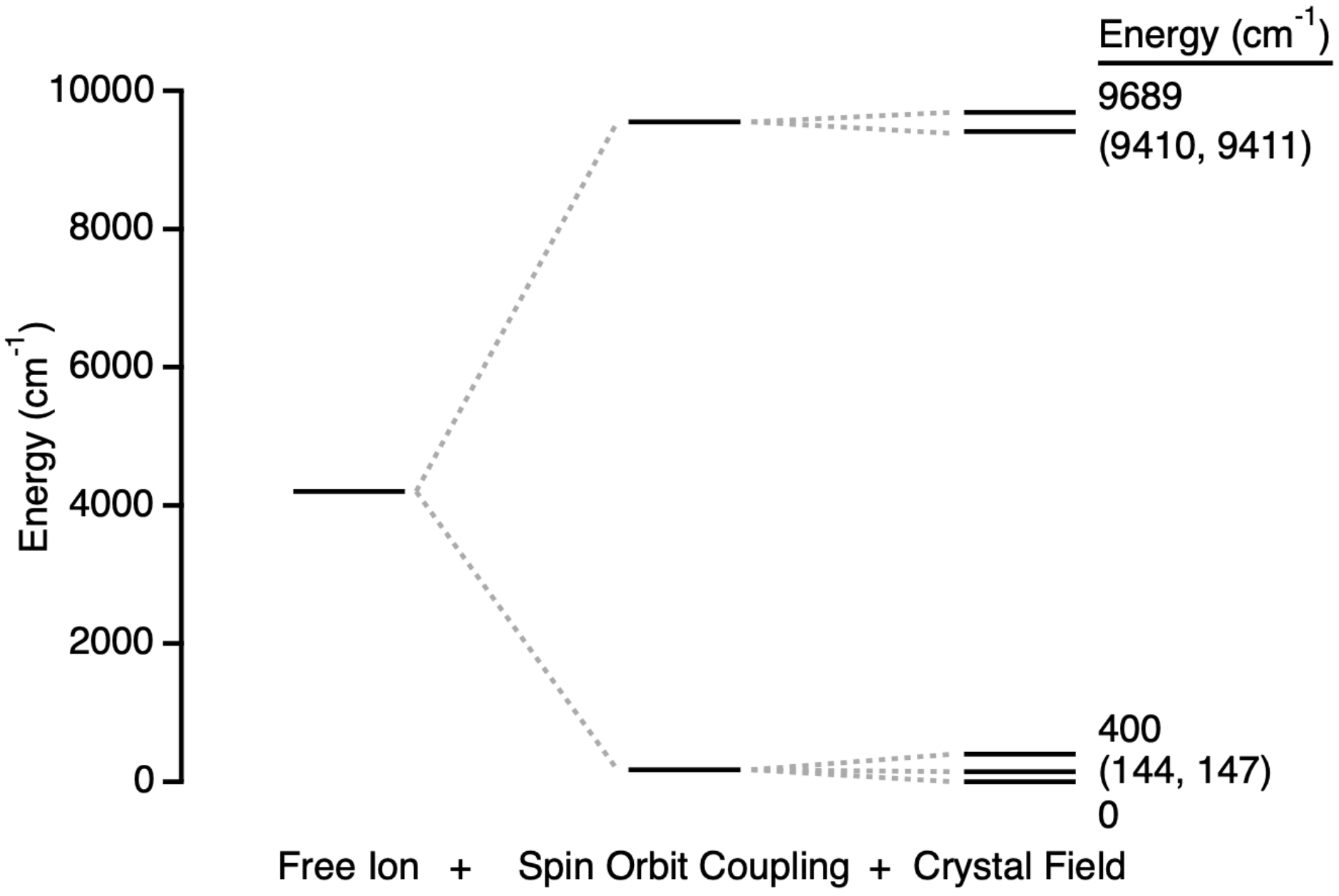

**Figure S6.** The $Yb^{3+}$ valence energy level diagram described by the best-fit parameters of Fig. S5. The energies of the crystal field states in eV are: 0.0000, (0.0179, 0.0182), 0.0496, (1.1667, 1.1668), 1.2013 eV.

**Table S2. Energies ($cm^{-1}$) of the valence electronic states, $^2F_{5/2}$ and $^2F_{7/2}$ barycenter energies,[a] and $\Delta E$(Barycenter) for $Yb^{3+}$ ions in several host crystals, and for the free ion. These data were used to generate Fig. S7 (after converting to eV). Many of these entries are compiled in ref. 16.**

| Host Lattice | 0 | 1 | 2 | 3 | $^2F_{7/2}$ Barycenter | 0' | 1' | 2' | $^2F_{5/2}$ Barycenter | $\Delta E$(Bary) | ref. |
|---|---|---|---|---|---|---|---|---|---|---|---|
| $Ca_2Ga_2SiO_7$ (CGS) | 0.0 | 300 | 490 | 970 | 440 | 10250 | 10570 | 11010 | 10610 | 10170 | 17 |
| $SrLaGa_3O_7$ (SLG) | 0.0 | 220 | 386 | 910 | 379 | 10190 | 10450 | 11025 | 10555 | 10176 | 17 |
| $Ca_4GdO(BO_3)_3$ (GdCOB) (site I, Gd) | 0.0 | 423 | 668 | 1003 | 524 | 10246 | 10706 | 11089 | 10680 | 10157 | 18 |
| GdCOB (site II, Ca) | 0.0 | 437 | 694 | 1003 | 534 | 10261 | 10737 | 11150 | 10716 | 10183 | 18 |
| GdCOB (site III, Ca) | 0.0 | 417 | 688 | 1003 | 527 | 10240 | 10682 | 11026 | 10649 | 10122 | 18 |
| $Ca_4YO(BO_3)_3$ (YCOB) | 0.0 | 427 | 556 | 1023 | 502 | 10242 | 10537 | 11109 | 10629 | 10128 | 19 |
| $Sc_2O_3$ | 0.0 | 474 | 634 | 1076 | 546 | 10250 | 10640 | 11198 | 10696 | 10150 | 20 |
| $Ca_5(PO_4)_3F$ (CFAP) | 0.0 | 409 | 597 | 1099 | 526 | 10178 | 10496 | 11069 | 10581 | 10055 | 21 |
| $Sr_5(PO_4)_3F$ (SFAP) | 0.0 | 362 | 618 | 1190 | 543 | 10150 | 10512 | 11108 | 10590 | 10048 | 22 |
| $Sr_5(VO_4)_3F$ (SVAP) | 0.0 | 321 | 562 | 1078 | 490 | 10141 | 10740 | 11050 | 10644 | 10153 | 23 |
| $Y_3Al_5O_{12}$ (YAG) | 0.0 | 584 | 635 | 783 | 501 | 10328 | 10752 | 10917 | 10666 | 10165 | 24 |
| $BaCaBO_3F$ (BCBF) | 0.0 | 303 | 533 | 902 | 435 | 10204 | 10570 | 11000 | 10591 | 10157 | 25 |
| $LiNbO_3$ | 0.0 | 352 | 448 | 788 | 397 | 10204 | 10471 | 11090 | 10588 | 10191 | 26 |
| $KGd(WO_4)_2$ (KGW) | 0.0 | 163 | 385 | 535 | 271 | 10188 | 10471 | 10682 | 10447 | 10176 | 27 |
| $KY(WO_4)_2$ (KYW) | 0.0 | 169 | 407 | 568 | 286 | 10187 | 10476 | 10695 | 10453 | 10167 | 27 |
| $CaWO_4$ | 0.0 | 220 | 366 | 492 | 270 | 10278 | 10366 | 10665 | 10436 | 10167 | 28 |
| $YAlO_3$ | 0.0 | 209 | 341 | 590 | 285 | 10220 | 10410 | 10730 | 10453 | 10168 | 28 |
| $LiYF_4$ | 0.0 | 216 | 371 | 479 | 267 | 10288 | 10409 | 10547 | 10415 | 10148 | 28 |
| $YAl_3(BO3)_4$ (YAB) | 0.0 | 94 | 185 | 581 | 215 | 10194 | 10277 | 10672 | 10381 | 10166 | 29 |
| $Cs_2NaYbCl_6$ | 0 | 225 | 225 | 573 | 256 | 10243 | 10243 | 10708 | 10398 | 10142 | 30, 31 |
| $Cs_3Yb_2Br_9$ | 0.0 | 144 | 201 | 421 | 192 | 10277 | 10301 | 10578 | 10385 | 10194 | 32 |
| $CsCdBr_3$ | 0.0 | 114 | 140 | 441 | 174 | 10119 | 10146 | 10590 | 10285 | 10111 | 32 |
| $CuInS_2$ | 0.0 | 32 | 87 | 182 | 75 | 10033 | 10060 | --- | 10095[a] | 10020 | 33 |
| InP | 0 | 35.5 | 35.5 | 97.5 | 42 | 10018 | 10064 | 10064 | 10049 | 10007 | 34 |
| Free ion | --- | --- | --- | --- | 0.0 | --- | --- | --- | 10213 | 10213 | 35 |
| $CrI_3$ | 0.0 | 146 | 146 | 400 | 173 | 9410 |  | --- | 9551[a] | 9379 | this work |

[a]For the entire data set of complete entries, the ratio of $^2F_{5/2}$:$^2F_{7/2}$ CF splitting energies, ($E$($^2F_{5/2}$ Barycenter) - $E_{0'}$)/($E$($^2F_{7/2}$ Barycenter)) is 0.82 ± 0.14. The $^2F_{5/2}$ barycenter energies for $Yb^{3+}$:$CrI_3$ and $Yb^{3+}$:$CuInS_2$ were thus set equal to the $^2F_{7/2}$ barycenter energies for the same compounds. The resulting uncertainties in $\Delta E$(Bary) are estimated to be < ~1%, close to or smaller than the data points in Fig. S7. For comparison, the $Yb^{3+}$:$CrI_3$ AOM calculations above yield: $^2F_{7/2}$ barycenter = 173 $cm^{-1}$ (21 meV), $^2F_{5/2}$ barycenter = 9503 $cm^{-1}$ (1.178 eV), $\Delta E$(Bary) = 9330 $cm^{-1}$ (1.157 eV), within this uncertainty range.

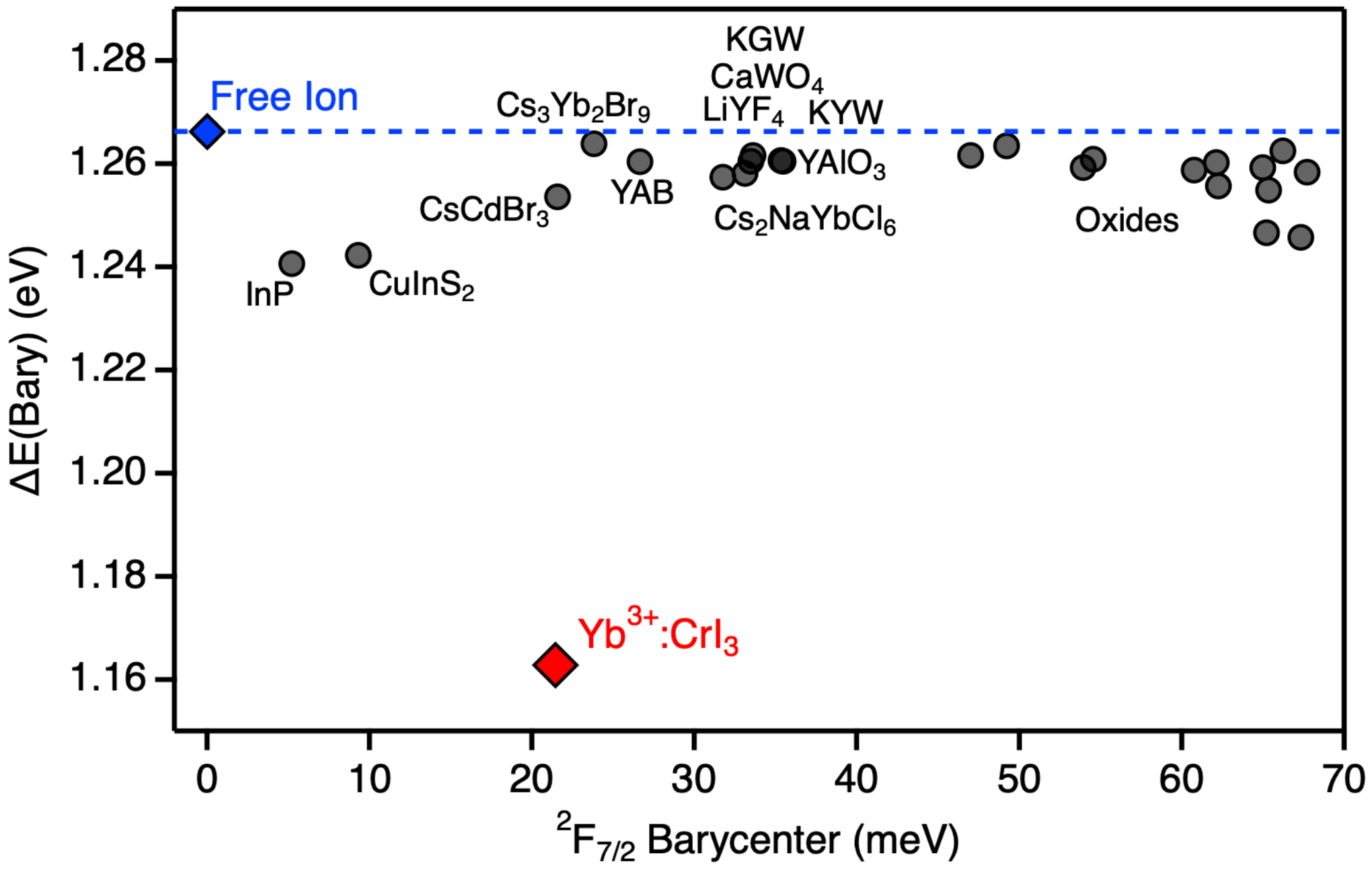

**Figure S7.** Plot of the difference between experimental $Yb^{3+}$ $^2F_{5/2}$ and $^2F_{7/2}$ barycenter energies ($\Delta E$(Bary)) for the compounds listed in Table S2, and for the free ion, *vs* the barycenter energy for the $^2F_{7/2}$ ground multiplet. The compounds associated with select data points are labeled. The dashed blue line shows the value of the free ion.

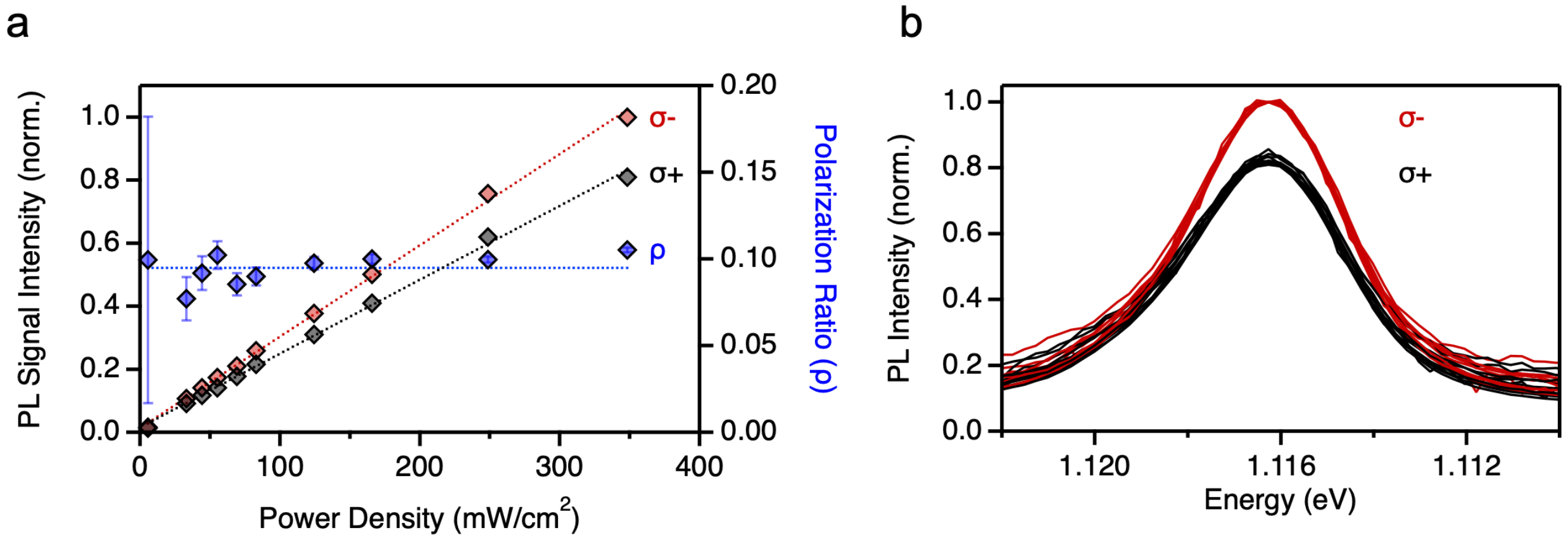

**Figure S8. (a)** Power dependence of $\sigma^-$ (red) and $\sigma^+$ (black) PL peak intensities and circular polarization ($\rho$, blue) of the $\Gamma_8 \rightarrow \Gamma_7$ transition. The data were collected at 0.5 T and 5 K and the sample was excited with linearly polarized light at 1.96 eV. The PL intensities show a linear increase with power, resulting in a constant polarization ratio. The error bars represent uncertainty estimated from the linear fit of the polarization intensities. **(b)** The $\sigma^-$ (red) and $\sigma^+$ (black) component of the $\Gamma_8 \rightarrow \Gamma_7$ transition normalized across all powers. The traces overlay each other well, showing no detectable power dependence.

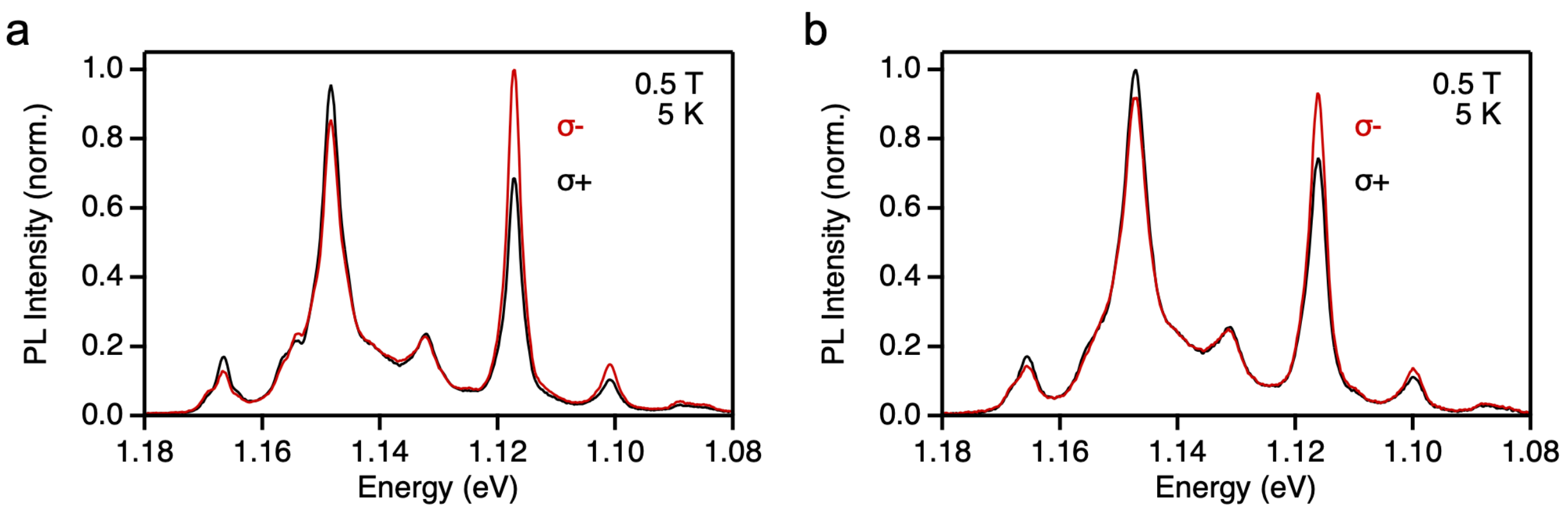

**Figure S9.** Comparison of full MCPL spectra across two different samples, measured at 0.5 T, 5 K. **(a)** The sample used in Fig. 3b,c,e,f of the main text. **(b)** The sample used in Fig. 3d of the main text. The two samples show very similar spectra, with slight differences in polarization magnitude.

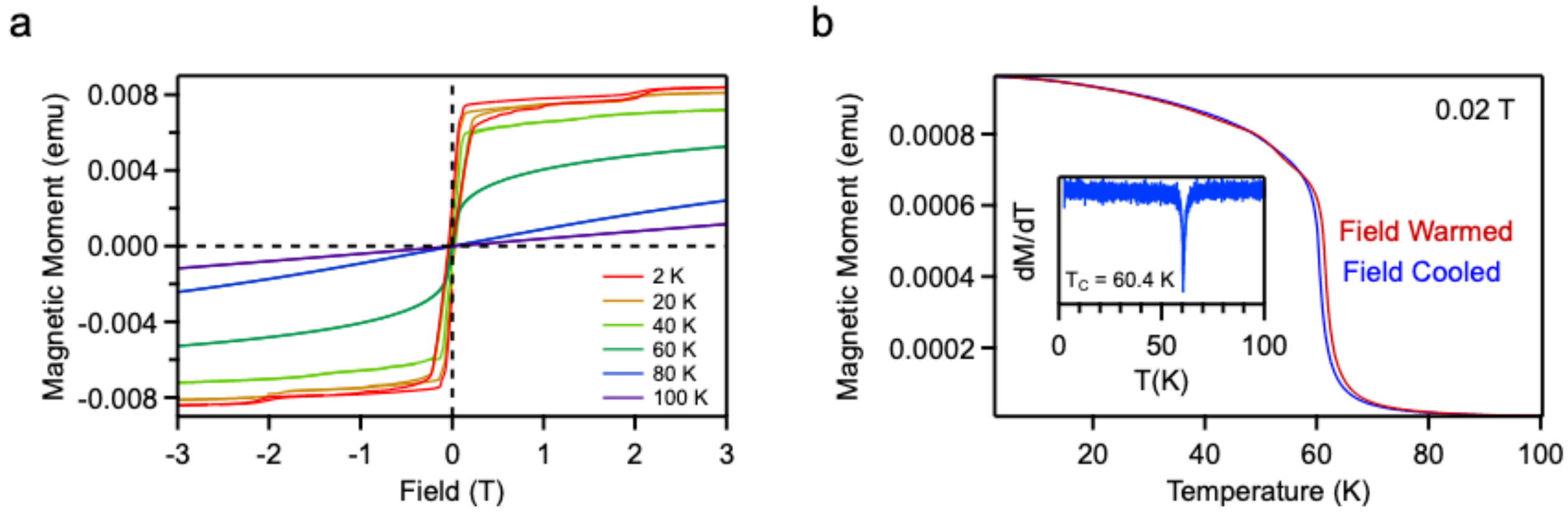

**Figure S10.** Magnetic data for a single-crystal flake of 5% $Yb^{3+}$:$CrI_3$, measured by VSM. The sample was probed with the external field aligned perpendicular to the face of the crystal. **(a)** Plots of magnetization vs external field measured at various temperatures. The data are similar to those collected on undoped $CrI_3$ bulk crystals (*e.g.*, Fig S11). At 2 K, a coercive field of ~44 mT was found. **(b)** Plot of magnetization *vs* temperature measured in the field-cooled and field-warmed directions. The inset shows the derivative of the field-cooled data as a function of temperature, where the Curie temperature is found to be 60.4 K. These data show that $Yb^{3+}$ doping has no significant effect on the magnetism of $CrI_3$ in these samples.

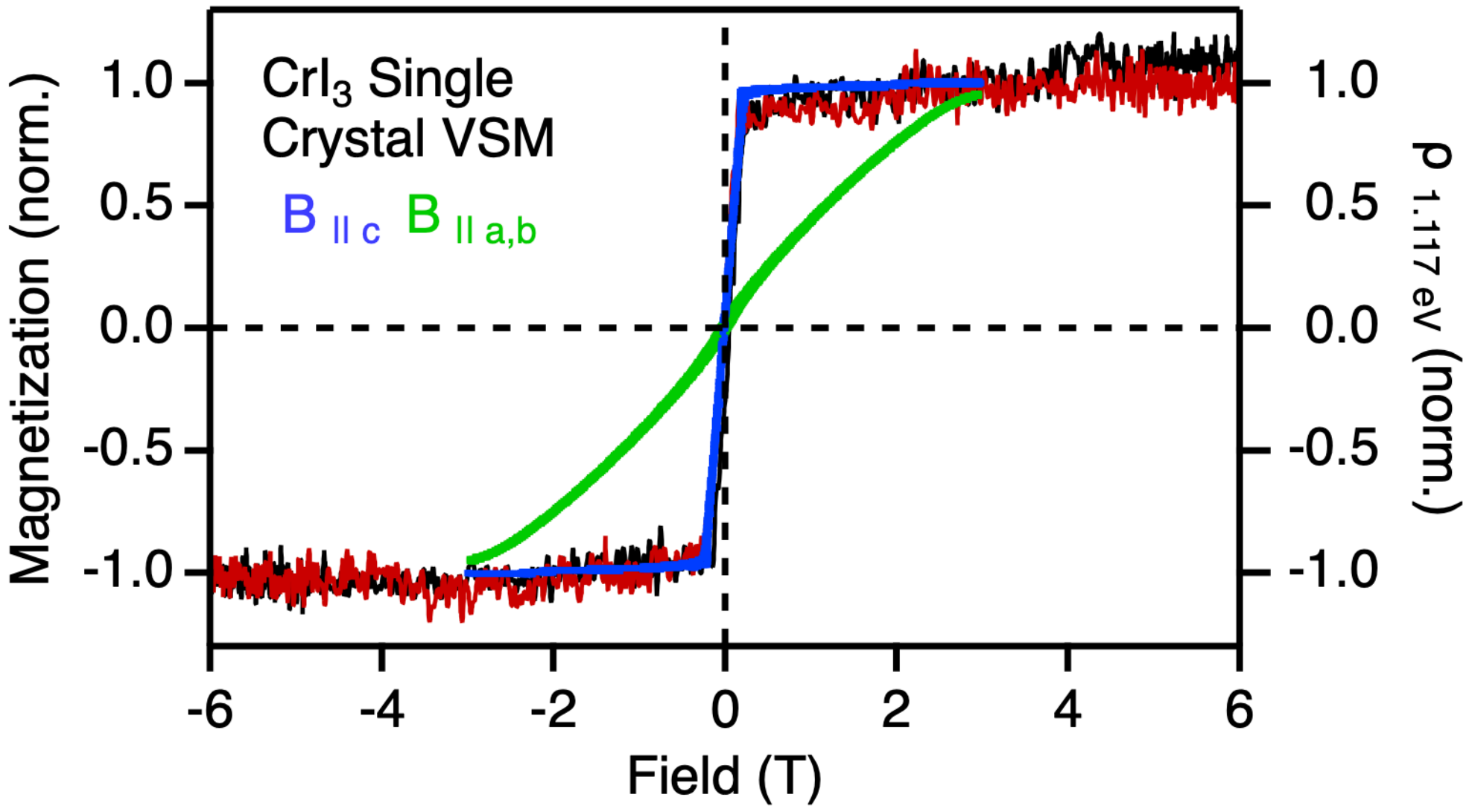

**Figure S11.** The same polarization data as featured in Fig. 3d of the main text, overlayed with $CrI_3$ magnetization data measured from -3 to +3 T with the field oriented parallel to the crystallographic *c* axis (blue) by single-crystal vibrating sample magnetometry (VSM).[36] For comparison, the magnetization perpendicular to *c* (green) is also shown. The $Yb^{3+}$ MCPL polarization $\rho$ is superimposable with the $CrI_3$ magnetization measured in the same configuration.

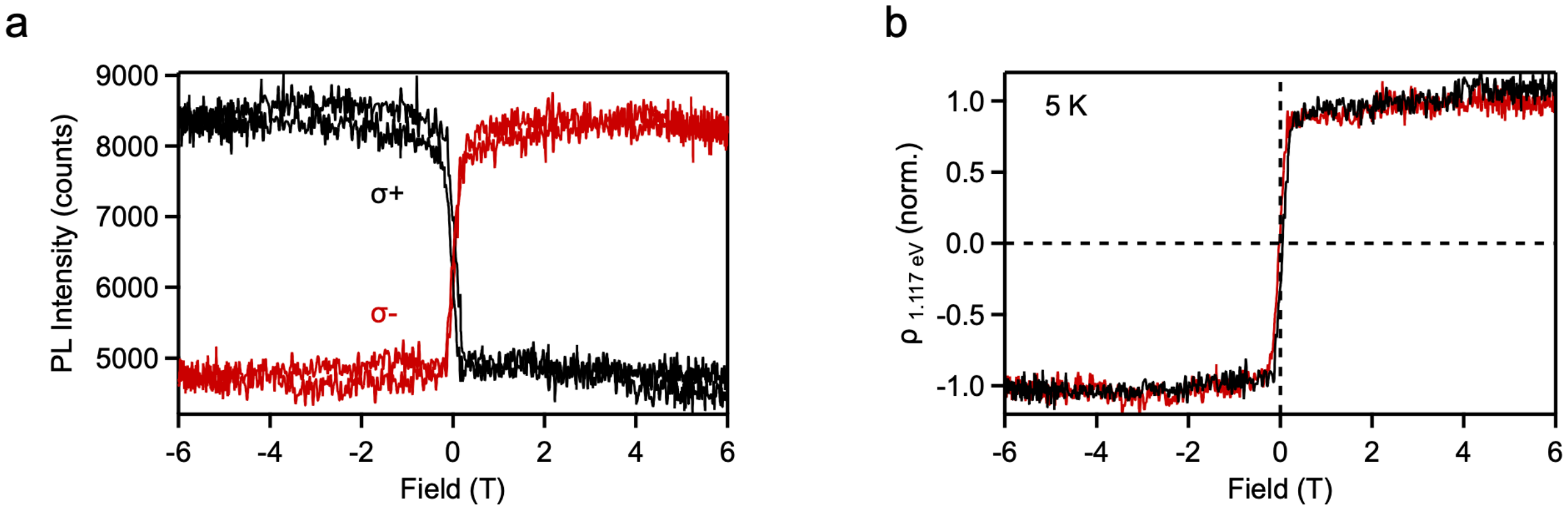

**Figure S12. (a)** Individual circularly polarized MCPL components measured during continuous field sweeps from -6 to +6 T and back at 5 K. **(b)** The same data, displayed as the polarization ratio ($\rho$, normalized). Panel (b) is shown as Fig. 3d of the main text. Data measured using 14 $mW/cm^2$ excitation.

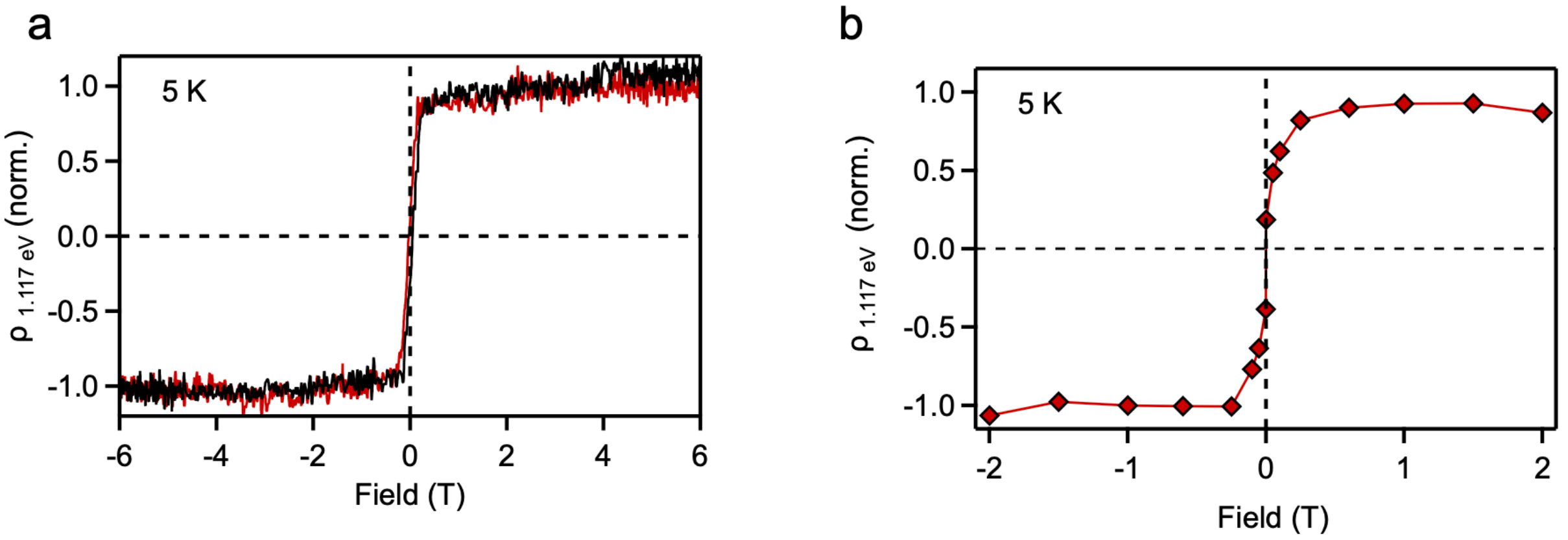

**Figure S13.** Comparison of field-dependent polarization ratios ($\rho$, normalized) measured with **(a)** linearly polarized and **(b)** unpolarized excitation at 5 K. In panel (b), no data were collected above 2 T. Panel (a) is shown as Fig. 3d of the main text.

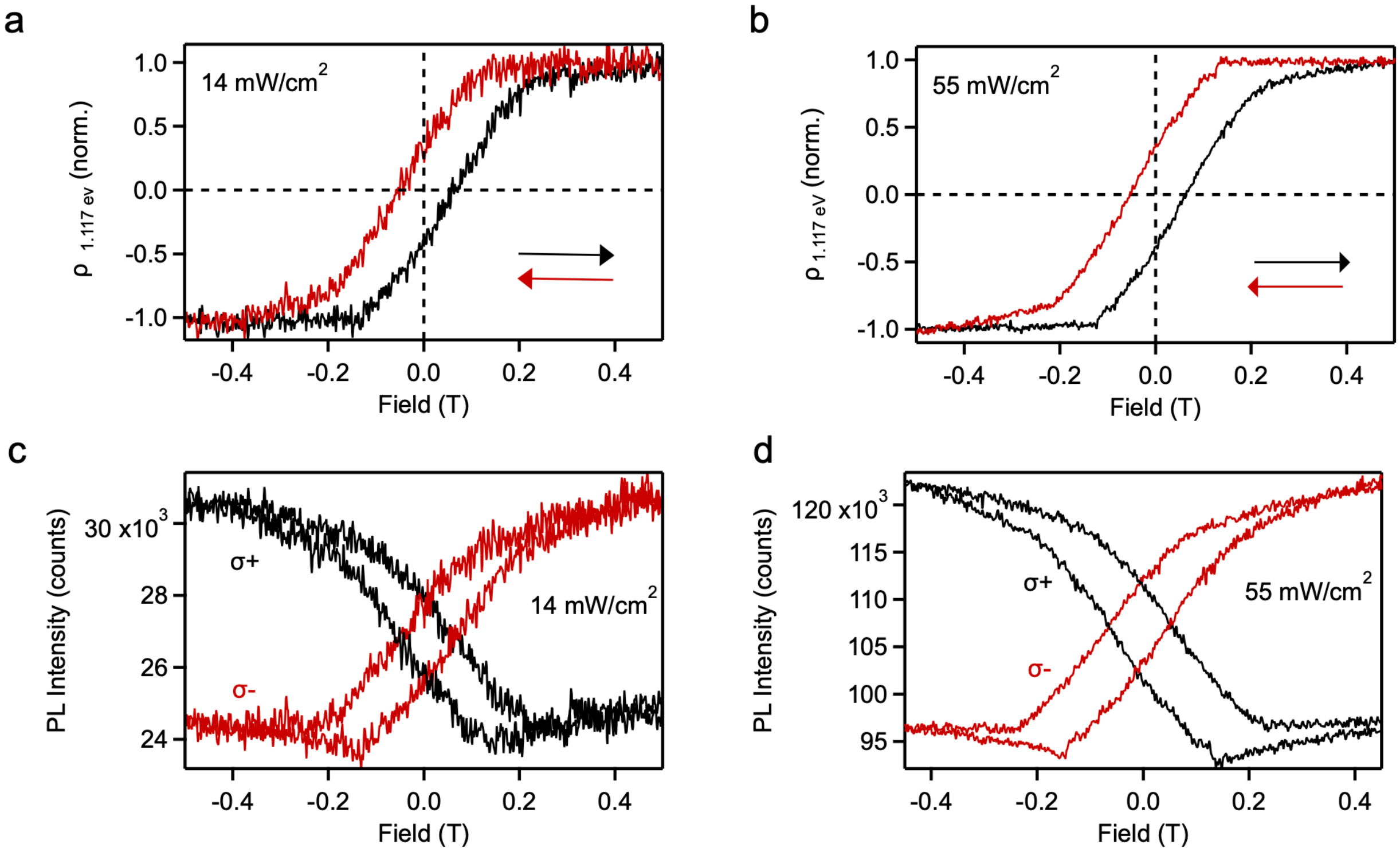

**Figure S14. (a,b)** Effect of excitation power on the polarization ratio ($\rho$, normalized). Magnetic hystereses measured under (a) low- and (b) higher-power excitation (14 *vs* 55 $mW/cm^2$, 5 K) show no difference. The black (red) trace corresponds to the sweep from negative (positive) to positive (negative) fields. **(c, d)** The separate circularly polarized PL components from the same (c) low- and (d) high-power measurements.

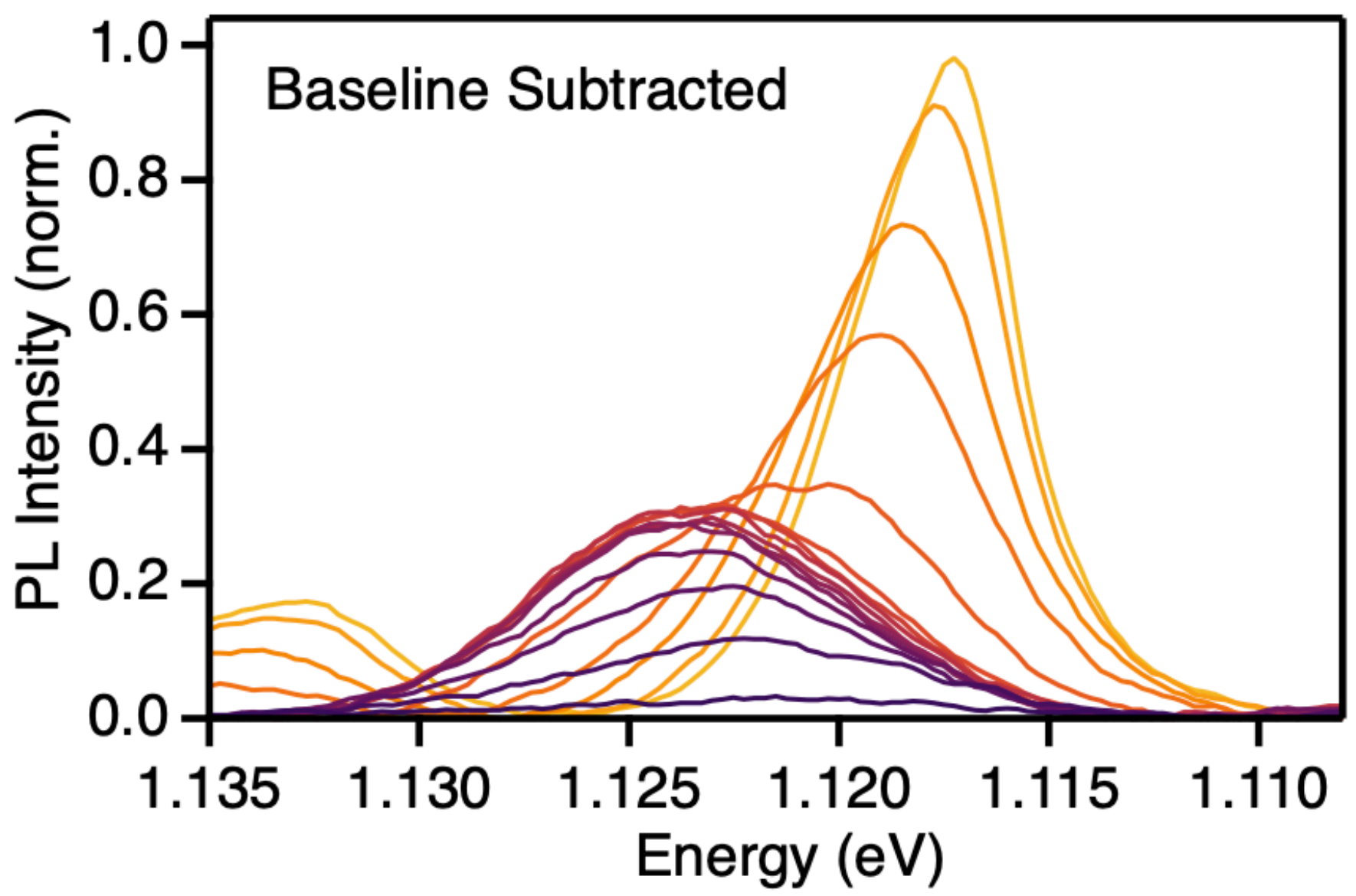

**Figure S15.** Temperature dependence of the $\Gamma_8 \rightarrow \Gamma_7$ PL feature of 4.9% $Yb^{3+}$:$CrI_3$ measured from 4 to 200 K under no external magnetic field (from Fig. 4 of the main text, T = 4, 15, 30, 40, 50, 55, 58, 60, 62, 65, 70, 85, 100, 125, 150 K). A linear baseline was subtracted from each spectrum here to facilitate viewing and determination of the peak's FWHM.

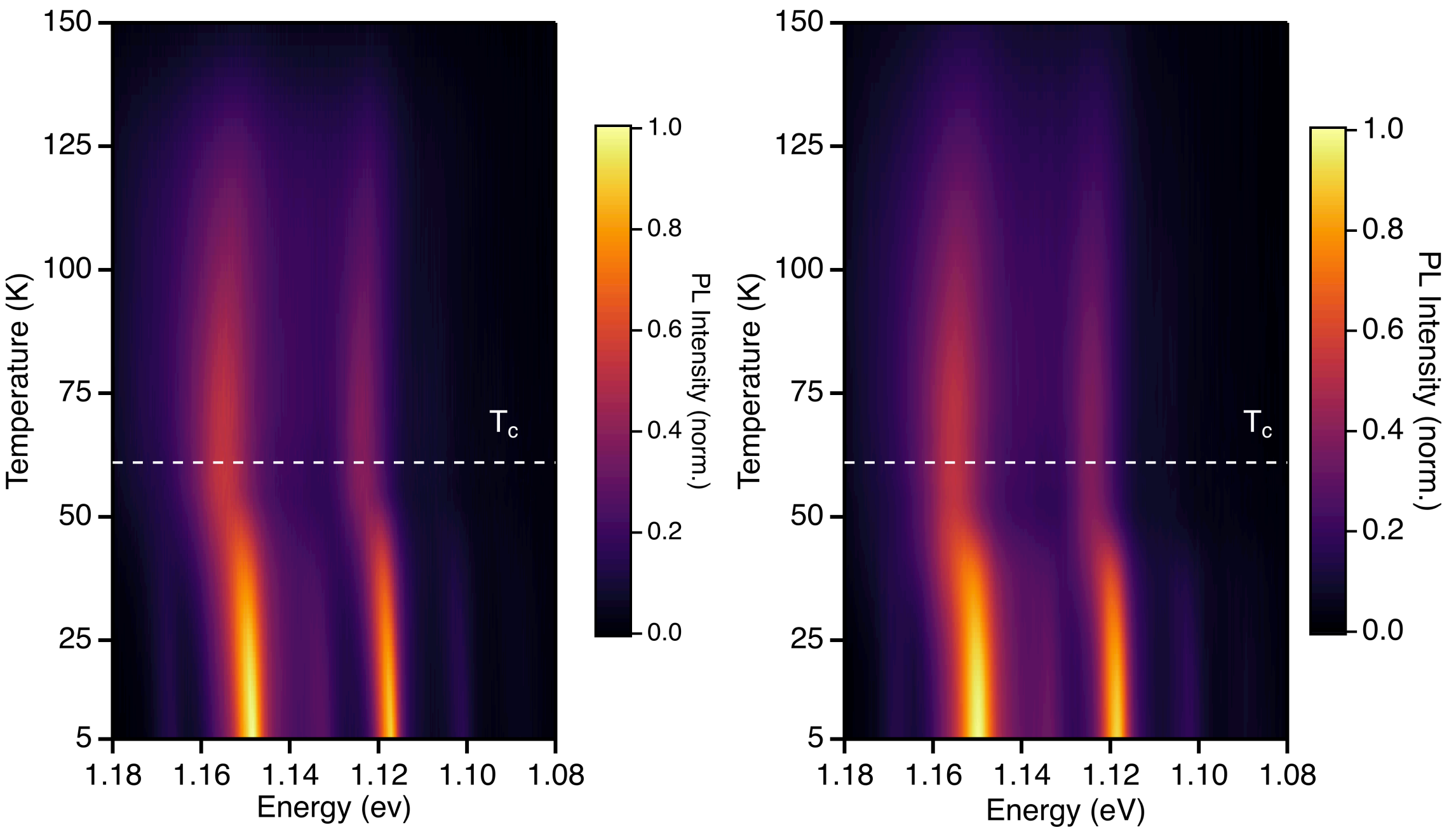

**Figure S16. (a,b)** False-color plots of the $Yb^{3+}$ PL intensities *vs* temperature measured for the two samples shown in Fig. S4c,e, respectively, from 4 to 150 K at zero external magnetic field. The horizontal dashed line indicates $T_C$ = 61 K. The two samples show the same temperature dependence, but the features are slightly better resolved in panel (a). Panel (a) is shown as Fig. 4a of the main text.

**References**

(1) Seyler, K. L.; Zhong, D.; Klein, D. R.; Gao, S.; Zhang, X.; Huang, B.; Navarro-Moratalla, E.; Yang, L.; Cobden, D. H.; McGuire, M. A.; Yao, W.; Xiao, D.; Jarillo-Herrero, P.; Xu, X. Ligand-field helical luminescence in a 2D ferromagnetic insulator. *Nat. Physics* **2018**, *14* (3), 277-281.

(2) Abramoff, M. D.; Magalhaes, P. J.; Ram, S. J. Image Processing with ImageJ. *Biophot. Int.* **2004**, *11* (7), 35-42.

(3) Bruker. APEX2 (Version 2.1-4), SAINT (version 7.34A), SADABS (version 2007/4),. **2007**,

(4) Sheldrick, G. M. A short history of SHELX. *Acta Cryst.* **2007**, *A64*, 112-122.

(5) Sheldrick, G. M. SHELXT - Integrated space-group and crystal-structure determination. *Acta Cryst.* **2015**, *A71*, 3-8.

(6) Altomare, A.; Cascarano, G. L.; Giacovazzo, C.; Guagliardi, A. Completion and refinement of crystal structures with SIR92. *J. Appl. Cryst.* **1993**, *26*, 343-350.

(7) Altomare, A.; Burla, C.; Camalli, M.; Cascarano, G. L.; Giacovazzo, C.; Guagliardi, A.; Moliterni, A. G. G.; Polidori, G.; Spagna, R.; Burla, C.; Camalli, M.; Cascarano, G. L.; Giacovazzo, C.; Guagliardi, A.; Moliterni, A. G. G.; Polidori, G.; Spagna, R. SIR97: a new tool for crystal structure determination and refinement. *J. Appl. Crystallogr.* **1999**, (32), 115-119.

(8) Sheldrick, G. M. SHELXL-97, Program for the Refinement of Crystal Structures. **1997**,

(9) Sheldrick, G. M. Crystal structure refinement with SHELXL. *Acta Cryst.* **2015**, *C71*, 3-8.

(10) Waasmaier, D.; Kirfel, A. New Analytical Scattering Factor Functions for Free Atoms and Ions. *Acta Crysta.* **1995**, *A51* (416-430),

(11) Piepho, S. B.; Schatz, P. N. *Group theory in spectroscopy with applications to magnetic circular dichroism*; John Wiley & Sons, 1983.

(12) Bronova, A.; Bredow, T.; Glaum, R.; Riley, M. J.; Urland, W. BonnMag: Computer program for ligand-field analysis of $f^n$ systems within the angular overlap model. *J. Comp. Chem.* **2018**, *39* (3), 176-186.

(13) McGuire, M. A.; Dixit, H.; Cooper, V. R.; Sales, B. C. Coupling of Crystal Structure and Magnetism in the Layered, Ferromagnetic Insulator $CrI_3$. *Chem. Mater.* **2015**, *27* (2), 612-620.

(14) Jesche, A. F., M.; Kreyssig, A.; Meier, W. R.; Canfield, P. C. X-Ray Diffraction on large single crystals using a powder diffractometer. *Philos, Mag. (Abingdon)* **2016**, *96* (20), 2115-2124.

(15) Kuindersma, S. R.; Boudewijn, P. R.; Haas, C. Near-Infrared d–d Transitions of $NiI_2$, $CdI_2$:$Ni^{2+}$, and $CoI_2$. *Phys. stat. sol. (b)* **1981**, *108* (1), 187-194.

(16) Haumesser, P.-H.; Gaumé, R.; Viana, B.; Antic-Fidancev, E.; Vivien, D. Spectroscopic and crystal-field analysis of new Yb-doped laser materials. *J. Phys.: Cond. Mat.* **2001**, *13* (23), 5427-5447.

(17) Simondi-Teisseire, B. PhD Thesis. Paris VI University, 1996.

(18) Mougel, F.; Dardenne, K.; Aka, G.; Kahn-Harari, A.; Vivien, D. Ytterbium-doped $Ca_4GdO(BO_3)_3$: an efficient infrared laser and self-frequency doubling crystal. *J. Opt. Soc. Am. B* **1999**, *16* (1), 164-172.

(19) Mougel, F. PhD Thesis. Paris VI University, 1999

(20) Mix, E. PhD Thesis. Hamburg University, 1999.

(21) DeLoach, L. D.; Payne, S. A.; Chase, L. L.; Smith, L. K.; Kway, W. L.; Krupke, W. F. Evaluation of absorption and emission properties of $Yb^{3+}$ doped crystals for laser applications. *IEEE J. Quant. Elect.* **1993**, *29* (4), 1179-1191.

(22) Gruber, J. B.; Zandi, B.; Merkle, L. Crystal-field splitting of energy levels of rare-earth ions $Dy^{3+}(4f^9)$ and $Yb^{3+}(4f^{13})$ in M (II) sites in the fluorapatite crystal $Sr_5(PO_4)_3F$. *J. Appl. Phys.* **1998**, *83* (2), 1009-1017.

(23) Payne, S. A.; DeLoach, L. D.; Smith, L. K.; Kway, W. L.; Tassano, J. B.; Krupke, W. F.; Chai, B. H. T.; Loutts, G. Ytterbium□doped apatite□structure crystals: A new class of laser materials. *J. Appl. Phys.* **1994**, *76* (1), 497-503.

(24) Bogomolova, G. A.; Bumagina, L. A.; Kaminskii, A. A.; Malkin, B. Z. Crystal field in laser garnets with $TR^{3+}$ ions in the exchange charge model. *Sov. Phys. Solid State* **1977**, *19* (8), 1428-1435.

(25) Schaffers, K. I.; DeLoach, L. D.; Payne, S. A. Crystal growth, frequency doubling, and infrared laser performance of $Yb^{3+}$: $BaCaBO_3F$. *IEEE J. Quant. Elect.* **1996**, *32* (5), 741-748.

(26) Montoya, E.; Sanz-Garcıa, J.; Capmany, J.; Bausá, L.; Diening, A.; Kellner, T.; Huber, G. Continuous wave infrared laser action, self-frequency doubling, and tunability of $Yb^{3+}$: MgO: $LiNbO_3$. *J. Appl. Phys.* **2000**, *87* (9), 4056-4062.

(27) Kuleshov, N. V.; Lagatsky, A. A.; Podlipensky, A. V.; Mikhailov, V. P.; Huber, G. Pulsed laser operation of Yb-doped $KY(WO_4)_2$ and $KGd(WO_4)_2$. *Optics lett.* **1997**, *22* (17), 1317-1319.

(28) Morrison, C. A.; Leavitt, P. Handbook on the physics and chemistry of rare earths, ch 46. Amsterdam: Elsevier: 1982.
(29) Wang, P.; Dawes, J. M.; Dekker, P.; Knowles, D. S.; Piper, J. A.; Lu, B. Growth and evaluation of ytterbium-doped yttrium aluminum borate as a potential self-doubling laser crystal. *J. Opt. Soc. Am. B* **1999**, *16* (1), 63-69.
(30) Schwartz, R. W. Electronic structure of the octahedral hexachloroytterbate ion. *Inorg. Chem.* **1977**, *16* (7), 1694-1698.
(31) Kanellakopulos, B.; Amberger, H. D.; Rosenbauer, G. G.; Fischer, R. D. Zur Elektronenstruktur hochsymmetrischer Verbindungen der Lanthanoiden und Actinoiden—V: Paramagnetische Suszeptibilität und elektronisches Raman-Spektrum von $Cs_2NaYb(III)Cl_6$. *J. Inorg. Nuc. Chem.* **1977**, *39* (4), 607-611.
(32) Malkin, B. Z.; Leushin, A. M.; Iskhakova, A. I.; Heber, J.; Altwein, M.; Moller, K.; Fazlizhanov, I. I.; Ulanov, V. A. EPR and optical spectra of $Yb^{3+}$ in $CsCdBr_3$: Charge-transfer effects on the energy-level structure of $Yb^{3+}$ in the symmetrical pair centers. *Phys. Rev. B* **2000**, *62* (11), 7063.
(33) Tsujii, N.; Imanaka, Y.; Takamasu, T.; Kitazawa, H.; Kido, G. Photoluminescence of $Yb^{3+}$-doped $CuInS_2$ crystals in magnetic fields. *J. Appl. Phys.* **2001**, *89* (5), 2706-2710.
(34) de Maat-Gersdorf, I. Spectroscopic analysis of erbium-doped silicon and ytterbium doped indium phosphide. University of Amsterdam, 2001.
(35) Wyart, J.-F.; Tchang-Brillet, W.-Ü. L.; Spector, N.; Palmeri, P.; Quinet, P.; Biémont, E. Extended Analysis of the Spectrum of Triply-ionized Ytterbium (Yb IV) and Transition Probabilities. *Phys. Scripta* **2001**, *63* (2), 113-121.
(36) De Siena, M. C.; Creutz, S. E.; Regan, A.; Malinowski, P.; Jiang, Q.; Kluherz, K. T.; Zhu, G.; Lin, Z.; De Yoreo, J. J.; Xu, X.; Chu, J.-H.; Gamelin, D. R. Two-Dimensional van der Waals Nanoplatelets with Robust Ferromagnetism. *Nano Lett.* **2020**, *20* (3), 2100-2106.